\documentclass{article} 

\AtBeginDocument{%
  \providecommand\BibTeX{{%
    \normalfont B\kern-0.5em{\scshape i\kern-0.25em b}\kern-0.8em\TeX}}}

\usepackage{color,soul}
\usepackage{algorithmicx}
 
\usepackage{algorithm}
\usepackage{algpseudocode}
\usepackage{adjustbox}
\usepackage[utf8]{inputenc}
\usepackage{url} 
\usepackage{hyperref} 
\usepackage{booktabs}
\usepackage{graphicx}
\usepackage{lscape}
\usepackage{wrapfig}
\usepackage{enumerate}
\usepackage{graphicx}
\usepackage{caption}
\usepackage{subcaption}
\usepackage{array}
\usepackage{tabularx}
\usepackage{wrapfig}
\usepackage{subcaption}
\usepackage{xcolor}
\usepackage{wrapfig}

\usepackage{pifont}
\usepackage{fullpage}

\begin{document}

\title{Exploring Heart Rate Variability and Heart Rate Dynamics Using Wearables Before, During, and After Speech Activity: Insights from a Controlled Study in a Low-Middle-Income Country}
\date{}

\author{Nilesh Kumar Sahu, Snehil Gupta, Haroon R. Lone}

\maketitle

\begin{abstract} 
\noindent
Conventional methods for diagnosing Social Anxiety Disorder (SAD), such as clinical interviews and self-reported questionnaires, often face accessibility barriers and subjective biases, underscoring the need for objective physiological markers. This study investigates heart rate (HR) and heart rate variability (HRV) as potential indicators of SAD by analyzing cardiovascular responses to anxiety-inducing speech tasks across four distinct phases: baseline, anticipation, speech activity, and reflection.
 
In a controlled laboratory setting, we analyzed data from 51 participants and found that HRV decreased and HR increased during the anticipation and speech activity phases compared to baseline, while the reflection phase showed a reversal, with HRV increasing and HR decreasing. Participants with SAD exhibited lower HRV, higher HR, and greater self-reported anxiety than non-SAD participants across all phases.
These findings enhance our understanding of the physiological signatures of social anxiety and have implications for developing wearable-based monitoring systems for SAD detection and intervention. To support further research, we also release a dataset capturing multi-phase anxiety responses, advancing physiological-based mental health assessment.\footnote{\textbf{This is the author's accepted manuscript of a paper accepted for publication in ACM COMPASS. 
The final published version is available via the ACM Digital Library at \url{<https://doi.org/10.1145/3715335.3735477 >}.}}.

\end{abstract}

\section{Introduction}

Social Anxiety Disorder (SAD), one of the anxiety spectrum disorders, is characterized by an overwhelming sense of fear and apprehension in social settings. SAD, also known as Social Phobia, significantly impairs one's ability to carry out daily tasks because of an intense fear of being negatively evaluated by others in various social situations, such as conversing with strangers or participating in classroom discussions. Consequently, individuals with SAD tend to avoid social situations due to the fear of embarrassment. As of 2020, around 19\% of adults in the US population (approx. 40 million individuals) are affected by the anxiety disorder~\cite{lecamwasam2023investigating}. Considering that anxiety ranks as one of the most prevalent mental health disorders worldwide, detecting and addressing it at a younger age is essential.


The lack of reliable and consistent physiological markers for SAD highlights the need to examine variations in various physiological parameters — such as heart rate, saliva composition, and blood pressure — among individuals in different anxious social situations. Since these parameters are controlled by the Autonomic Nervous System (ANS), which is sensitive to anxiety-inducing stimuli~\cite{lecamwasam2023investigating}, it is crucial to explore how they vary under different conditions and types of anxiety.
 Among the various physiological makers, Heart Rate (HR) and Heart Rate Variability (HRV) can be easily monitored nowadays as the wearables come with pre-installed Photoplethysmography (PPG) sensor.
However, research on HR and HRV variability during anxiety-inducing activities, such as public speaking or cognitive tasks, shows mixed results. Some studies report ``significantly reduced'' HRV~\cite{alvares2013reduced,gaebler2013heart,garcia2017autonomic}, while others report ``insignificantly reduced or increased'' HRV~\cite{held2021heart,pittig2013heart,tolin2021psychophysiological, cheng2022heart} during anxiety provoking situations in SAD individuals. 
Notably, Held et al.~\cite{held2021heart} report that these mixed outcomes are not necessarily due to the nature of anxiety-inducing activities but rather stem from the differing characteristics of participants. These variations in participant characteristics raise concerns about the global generalizability of existing research findings. Therefore, it becomes imperative to re-evaluate the relationship between cardiovascular measures (i.e., HR, HRV) and anxiety-provoking activities across different world regions.
 
In this paper, we introduce a novel study that examines for the first time how HR and HRV fluctuate over time—specifically before, during, and after an anxiety-provoking activity—within a low-middle-income country (LMIC). Our approach is innovative because it separately analyzes HR and HRV variations across three distinct phases:   Anticipation, Activity, and Reflection compared to the Baseline. This contrasts with the majority of existing research, which typically compares HR/HRV variations only at a single time point or immediately after an anxiety-inducing event~\cite{alvares2013reduced, miranda2014anxiety, tamura2013salivary, madison2021social}.

Our study makes an important contribution to affective computing, focusing on the Global South—a region that faces unique and often overlooked challenges in mental health. To our knowledge, this is the first three-phase study that explores the use of physiological markers for identifying mental health disorders in India.  In LMICs such as India, mental health issues are shaped by complex, intersecting factors such as social stigma, limited access to care, and cultural perceptions of illness \cite{pendse2019mental}. These factors deeply shape both the experience and treatment of mental health conditions, often exacerbating the suffering of affected individuals~\cite{patel2018lancet}.

\begin{figure*}[!t]
\centering
  \includegraphics[scale=0.50]{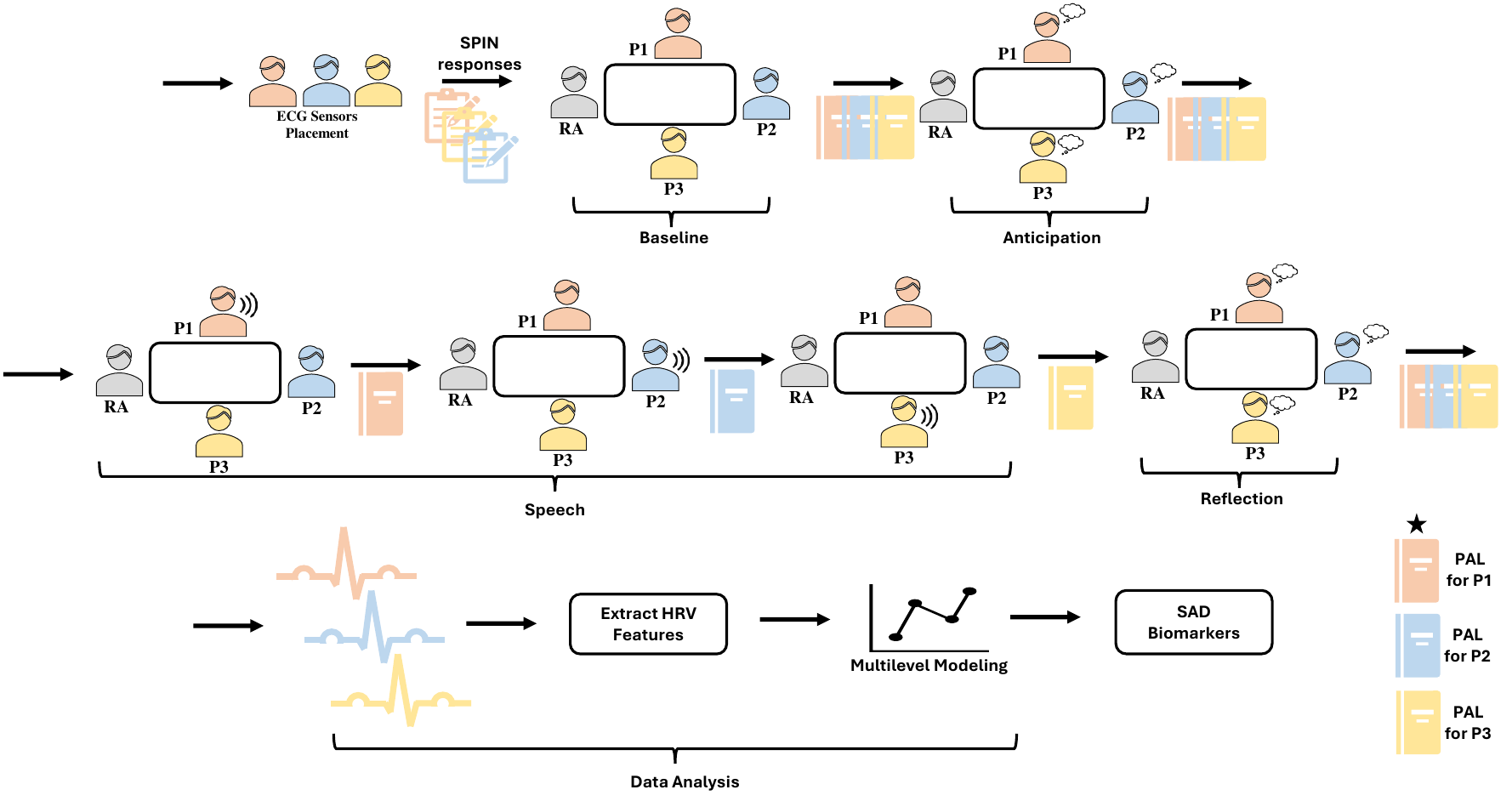}
  \caption{
  Each study session involved a Research Associate (RA) and three participants, labeled P1, P2, and P3, represented by pink, blue, and yellow colors. Initially, ECG sensors were attached to the participants, and they provided self-reported SPIN responses. The participants then proceeded through four phases: Baseline, Anticipation, Speech Activity, and Reflection. After the Baseline phase and each subsequent phase, participants reported their perceived anxiety levels (PAL) using a single-item questionnaire. For data analysis, HR and HRV features—covering time-domain, frequency-domain, and non-linear metrics—were extracted from cleaned ECG signals and analyzed using a multilevel model to determine their potential as physiological markers for SAD.
}
\label{fig: study_framework}
\end{figure*}

We conducted a controlled lab study involving ninety-nine non-clinical university students who underwent an anxiety-inducing task.  Participants were assessed during the Baseline phase and across three additional phases—Anticipation, Activity, and Reflection. We classified participants into SAD and non-SAD groups based on their Social Phobia Inventory (SPIN) scores, a standard measure for SAD severity. Figure~\ref{fig: study_framework} provides a high-level overview of our study. To test our following hypotheses, we employed hierarchical linear modeling (HLM), adding a novel methodological dimension to the analysis of anxiety-related physiological responses.
 
\begin{itemize}
 \item   \textbf{Hypothesis 1 (H1):} There will be a significant increase in HR and a decrease in HRV during the Anticipation and  Activity phases, respectively, compared to the Baseline for all participants. Additionally, HR and HRV will exhibit significant differences during the Reflection phase compared to the Baseline phase for all participants. This hypothesis does not consider grouping participants based on different attributes (e.g., gender, age) and relies solely on HR and HRV analysis.
 
  \item   \textbf{Hypothesis 2 (H2):} There will be significant differences in HR and HRV between the SAD and non-SAD participants during the Activity phase. This hypothesis considers grouping participants based on their anxiety severity (i.e., SAD, Non-SAD).
 
  \item  \textbf{Hypothesis 3 (H3):} There will be a significant increase in perceived anxiety levels (self-reported) during the Anticipation and Activity phases compared to the Baseline for all participants. Furthermore, perceived anxiety levels during the Reflection phase will significantly differ from those during the Baseline. This hypothesis differs from the previous two hypotheses as it investigates ``perceived anxiety" instead of HR or HRV. 
\end{itemize}

From our hypotheses testing, we have determined significant patterns in the participants' physiological responses. Specifically, we observed reduced  HRV during the anticipation and activity phases compared to the baseline, while HRV increased during the reflection phase for all participants. Moreover, we observed that SAD participants exhibited lower HRV compared to non-SAD. Our analysis revealed that HRV parameters such as RMSSD\footnote{The square root of the mean of squared successive differences between adjacent RR intervals}, SD1\footnote{The standard deviation perpendicular to the identity line, reflecting short-term RR interval fluctuations, i.e., beat-to-beat variability}, S\footnote{The area of an ellipse described by SD1 and SD2, proportional to SD1SD2}, and SDNN\footnote{The standard deviation of RR intervals} can discriminate between participants with SAD and those without it. Our examination of perceived anxiety levels indicated that all participants experienced heightened anxiety during both the anticipation and activity phases compared to the baseline. However, anxiety levels decreased during the reflection phase, highlighting a dynamic interplay between physiological responses and subjective emotional experiences across different study phases. Again, we found that SAD participants exhibited higher perceived anxiety levels compared to non-SAD. Following are the contributions of this paper.

\begin{itemize}
 
\item We design and implement novel three-phase study to investigate changes in the HR, HRV, and perceived anxiety during anticipation, activity, and reflection phases, separately. Unlike prior work, we analyze each phase separately to capture temporal dynamics of anxiety.

\item   We compile and publicly release a unique, high-resolution dataset\footnote{\url{https://osf.io/phze7/}} from a controlled laboratory study on SAD conducted in India. To the best of our knowledge, this is the first publicly available HRV dataset focused on SAD in a Global South context.

\item Our study contributes a culturally grounded understanding of anxiety in the Global South~\cite{pendse2019mental}. Cultural factors significantly shape how anxiety is experienced and expressed~\cite{vaillant2012positive, eshun2009introduction}. Existing literature~\cite{fernando2019developing,salamanca2020ethical,de2015missed,asher2017little} underscores the need for region-specific mental health research. To our knowledge, this is the first controlled study of SAD conducted in Indian setting.

\end{itemize}


\section{Literature Review} \label{sec: Literature_Review}

SAD has attracted attention from various academic disciplines, including computer science, physiology, psychology, and psychiatry, resulting in a diverse body of literature. In the context of our research, we have categorized related works into three distinct sections:
(i) \textit{Study designs:} This section classifies existing research examining cardio-vascular measures of SAD based on their study designs, i.e.,  distinguishing between single-phase and multiple-phase studies. (ii) \textit{Anxiety-inducing activities:} This section sheds light on the various anxiety-inducing activities employed across different studies to investigate SAD.
(iii) \textit{Wearables and smartphones for mental health disorder detection:} This section discusses the role of wearables and smartphones in controlling mental health disorders. The details of each of the mentioned sections are below.

\subsection{Study designs}

Research shows that individuals suffering from SAD may exhibit cardiovascular responses that are different from those without the disorder \cite{lecamwasam2023investigating,held2021heart}. HR and HRV are the most commonly measured cardiovascular physiological responses. HR measures the number of heartbeats per minute, while HRV quantifies the variations in time intervals between heartbeats, offering insights into how the ANS adapts to various stimuli \cite{havard_hrv, held2021heart}. Based on the study design, existing SAD studies can be categorized as single-phase or multiple-phase, as described below.
\\
 
\noindent\textit{i. Single-phase studies:} 
Such studies collect Electrocardiogram (ECG) data at rest (i.e., baseline) or during an anxious activity and study HR and HRV variations of SAD and non-SAD participants.
For example, Alavares et al. \cite{alvares2013reduced}, and Gaebler et al.~\cite{gaebler2013heart}  assessed HR, HRV at rest and found significantly reduced High Frequency, an HRV feature, in SAD participants compared to non-SAD. Additionally, Alavares et al.~\cite{alvares2013reduced} reported significant reductions in RMSSD and PCSD1\footnote{Poincaré plot perpendicular to the line of identity}, along with a significant increase in DFA$\alpha$1\footnote{The monofractal detrended fluctuation analysis of the HR signals} and mean HR in SAD participants. 
However, Gaebler et al.~\cite{gaebler2013heart} did not find any significant increase in HR but found an elevated LF/HF ratio and reduced HF during emotional face-matching tasks in SAD participants.

Similarly, Miranda et al. \cite{miranda2014anxiety} and Tamura et al. \cite{tamura2013salivary} analyzed HR and HRV during a presentation and electrical stimulation, respectively, and did not observe any significant increase in HR among SAD participants. Furthermore, Tamura et al. \cite{tamura2013salivary} reported no significant differences in the frequency domain HRV parameters (LF, HF, LF/HF) between SAD and non-SAD participants.

\noindent\textit{ii. Multiple-phase studies:} 
These studies investigate dynamic changes in HR and HRV before, during, and after engaging in anxiety-inducing activity. They provide a comprehensive view of how these parameters fluctuate in response to anxiety. For example, Tolin et al. \cite{tolin2021psychophysiological} and Rubio et al. \cite{garcia2017autonomic} examined changes in HRV across the baseline, Trier Social Stress Test (TSST) activity, and the recovery phase. However, the findings from these studies were different, i.e., Rubio et al. \cite{garcia2017autonomic} observed that SAD participants show significantly reduced RMSSD, while Tolin et al. \cite{tolin2021psychophysiological} did not identify any significant changes in RMSSD. Nevertheless, both studies reported a common finding: SAD participants exhibited higher HR. 

The HR and HRV findings vary across studies where some studies report ``significantly reduced'' HRV~\cite{alvares2013reduced,gaebler2013heart,garcia2017autonomic}, while others report ``insignificantly reduced or increased'' HRV~\cite{held2021heart,pittig2013heart,tolin2021psychophysiological, cheng2022heart} in SAD participants. Similarly, most \cite{miranda2014anxiety, tamura2013salivary, gaebler2013heart} studies did not find increased HR, whereas Alavares et al.\cite{alvares2013reduced} found Higher HR in SAD participants. The mixed findings necessitate further investigation to comprehensively understand the role of HRV and HR as potential physiological markers for SAD.

Our proposed study offers a unique perspective by examining how HR and HRV relate to an anxiety-inducing activity across different phases. Unlike existing research, which focussed on isolated aspects, we explore HR and HRV changes during three specific phases: anticipation, the activity itself, and reflection. This comprehensive approach is crucial for developing effective interventions. By understanding how anxiety impacts cardiovascular responses at each stage, we can tailor interventions to address better the severity and duration of anxiety experienced at different times during the activity.


\subsection{Anxiety-inducing activities}

In controlled settings, often anxiety-inducing activities are used to understand the effect of an anxious activity on different physiological (e.g., ECG) and psychological responses of participants.  These activities encompass a wide range of tasks, including speech~\cite{harrewijn2018heart, garcia2017autonomic, grossman2001gender,sahu2024unveiling}, hyperventilation~\cite{pittig2013heart}, cognitive challenges~\cite{held2021heart, gaebler2013heart, tolin2021psychophysiological, garcia2017autonomic, madison2021social}, reading~\cite{garcia2017autonomic}, presentations~\cite{miranda2014anxiety}, and behavioral assessment tests such as opposite-sex interaction~\cite{asher2020dating, asher2020out}. Table~\ref{tab:literature_review} lists various anxiety-inducing activities employed in previous studies.  It is worth noting that the duration of these anxiety-inducing activities varies across studies, reflecting the diverse approaches used to explore the physiological and psychological responses of participants with SAD in different contexts.

\subsection{Wearables and smartphones for mental disorder detection} 


Over the past decade, research has increasingly explored the use of wearables~\cite{xue2022understanding,lecamwasam2023investigating} and smartphones~\cite{salekin2018weakly, rashid2020predicting, wang2023detecting} for detecting physiological and behavioral markers of mental disorders. For instance, Lecamwasam et al.~\cite{lecamwasam2023investigating} employed Empatica E4 sensors to monitor participant's HR and Electrodermal activity (EDA) during mathematical problem-solving tasks, accompanied by personalized music to examine its impact on stress and anxiety induced by the tasks. Likewise, Xue et al.~\cite{xue2022understanding} used chest-mounted ECG heart rate sensors and explored how group members collectively reflected on organizational stress using a shared and anonymous visualization of HRV.
Notably, most studies involving wearables have primarily focused on stress~\cite{wang2022first, adler2021identifying, mishra2020evaluating, yu2023semi, xue2022understanding, lecamwasam2023investigating}.

Although the symptoms of mental disorders are consistent globally, their manifestations and underlying causes can differ due to societal and cultural variations~\cite{meegahapola2023generalization}. 
Thus, it is important to conduct studies in diverse geographical locations to uncover the unique causes of anxiety in different regions. In this context, our proposed wearable-based controlled study is a pioneering effort from a lower-middle-income country to identify potential physiological markers of SAD. To our knowledge, most wearable-based studies on anxiety have been conducted in developed countries. We believe our findings will complement and extend the insights gained from research in developed nations.

\begin{table}[!h]
\caption{Summary of SAD studies studying variations in HR and HRV features during different anxious activities.} 
\tiny
\centering

\hfill \break
\begin{tabular}{ p{0.2cm} >{\centering}p{2.5cm} >{\centering}p{0.3cm} >{\centering}p{0.5cm} >{\centering}p{2cm} >{\centering}p{2cm} >{\centering}p{1cm} p{3.9cm}}
 \toprule
 \textbf{Ref.} & \textbf{Objective} & \textbf{SAD} & \textbf{Control} & \textbf{Experimental phases} & \textbf{HRV} & \textbf{Analysis} & \textbf{Key Finding (s)} \\
 & & \textbf{(\#)} & \textbf{(\#)} & \textbf{(Activities)} & \textbf{measures} & \textbf{approach} &  \\
 \midrule
\cite{alvares2013reduced} &
Examine HRV in clinical SAD patients &
53 &
53 &
Resting (5 min) &
SDNN, RMSSD, HF, LF, PCSD1, DFA$\alpha$1 &
Between group &
Significantly increased HR and DFA$\alpha$1 and significantly reduced SDNN, RMSSD, and PCSD1
\\
\addlinespace[0.1cm]
\cite{miranda2014anxiety} &
Use wearables to detect anxiety level &
4 &
8 &
Presentation (10 min) &
HR &
Between group&
Insignificant HR increase in the mild SAD group 
\\
\addlinespace[0.2cm]
\cite{gaebler2013heart} &
Investigate HRV in SAD patients &
22 &
22 &
Resting (5min) and Emotional face matching task (approx. 5min) &
HR, HF, LF/HF &
Between group &
Rest: Significantly reduced HF and increased LF/HF ratio in SAD group but no difference in HR.
Task: Significantly less HF in SAD group.
\\
\addlinespace[0.2cm]
\cite{tolin2021psychophysiological} &
Examine psychophysiological arousal variations in different phases &
16 &
28 &
Baseline (3 min), stressor (6 min), Recovery (3 min) &
HF, RMSSD, HR &
Mixed linear model &
Increased HR in SAD compared to control.
Reduced HF in SAD during stressor. No significant changes in RMSSD 
\\
\addlinespace[0.2cm]
\cite{garcia2017autonomic} &
Study how SAD individuals react before, during, and after the TSST activity &
18 &
21 &
Baseline (10 min), TSST activity (5 min), Recovery (35 min) &
HR, RMSSD, LF/HF &
Both between and within &
Significantly higher HR and lower RMSSD in SAD individuals compared to controls.
\\
\addlinespace[0.2cm]
\cite{tamura2013salivary} &
Compare HRV of SAD individuals with controls &
32 &
80 &
Stimulation (40s) &
LF, HF, LF/HF, HR &
Between &
No significant difference in HR and HRV between SAD and non-SAD groups.
\\
\addlinespace[0.2cm]
\cite{bailey2019moderating} &
Examine HRV in SAD individuals in a 24-hour study &
16 &
16 &
24hr (Lying, Sitting, Standing, Walking, Running, Wake, Sleep) &
RMSSD, HR &
General linear mixed modeling &
No significant difference between HR and HRV during the wake. Further, they found higher HR during sleep. However no group difference in RMSSD during sleep. 
\\
\addlinespace[0.2cm]
\cite{harrewijn2018heart} &
Investigate whether HRV during rest and social tasks is a candidate endophenotype of SAD &
17 &
104 &
Resting state (3 min), anticipation  (5 min), speech  (3 min), recovery  (5 min), and second resting state  (5 min) &
RMSSD, HF, LF/HF, HR &
Regression model &
Participants with SAD exhibited lower HRV compared to controls. However, the results were not significant.
\\
\addlinespace[0.2cm]
\cite{madison2021social} &
Examine association between SAD symptoms and HRV &
124 &
- &
Resting state (5 min), Vocal emotion recognition (2 min) &
RMSSD &
Linear regression &
Individuals with greater social anxiety symptoms have significantly lower task HRV but not lower resting HRV. 
\\
\addlinespace[0.2cm]
\cite{grossman2001gender} &
Examine cardiovascular responses in older people in social situations &
30 &
30 &
Baseline (10min), paced breathing (5 min), speech preparation (4 min), speech (4 min) &
HR &
MANOVA&
Significantly increased  HR for both the groups from baseline to speech preparation and speech presentation. Women had significantly increased HR than men in both tasks.
 \\
\addlinespace[0.2cm]
\cite{pittig2013heart} &
Examine HF-HRV in SAD and other anxiety disorder participants &
25 &
39 &
Baseline (5 min), relaxation (15 min), hyperventilation (1 min) &
HR and HF &
Hierarchical linear modeling &
No significant difference in HR in baseline, relaxation, and hyperventilation between SAD and non-SAD. However, non-SAD groups show significantly higher HF in the baseline, with no difference in relaxation and hyperventilation.
\\
\addlinespace[0.2cm]
\cite{held2021heart} &
Investigate HRV and HR in clinically anxious participants
&
26&
14&
Baseline (3 min), Working memory Task (10 min), Recovery (3 min)&
HF and HR&
Multilevel modeling &
No significant difference in HF and HRV at baseline and reflection phase.
\\
\addlinespace[0.2cm]
\bottomrule
\end{tabular}

\label{tab:literature_review}
\end{table}


\section{Methodology} \label{sec: Methodology}

\begin{table}
\centering
\caption{Participants characteristics}
\label{tab:participant_characterstics}
\begin{tabular}{@{}lccccc@{}}
\toprule
 & \multicolumn{1}{c}{\begin{tabular}[c]{@{}c@{}} \textbf{Participants} \\ \textbf{(\#)}\end{tabular}} & 
 \multicolumn{1}{c}{\begin{tabular}[c]{@{}c@{}}\textbf{SPIN score}\\ \textbf{($\mu$, $\sigma$)}\end{tabular}} & 
 \multicolumn{1}{c}{\begin{tabular}[c]{@{}c@{}}\textbf{Age}\\ \textbf{($\mu$, $\sigma$)}\end{tabular}} & 
 \multicolumn{1}{c}{\begin{tabular}[c]{@{}c@{}}\textbf{Gender}\\ \textbf{(M, F)}\end{tabular}} & 
 \multicolumn{1}{c}{\begin{tabular}[c]{@{}c@{}}\textbf{Home location}\\ \textbf{(urban, rural)}\end{tabular}} \\ \midrule
All & 51 & (24.06, 11.20) & (21.10, 2.59) & (34, 17) & (37, 14) \\
SAD & 31 & (31.13, 7.86) & (20.48, 2.22) & (23, 8) & (24, 7) \\
non-SAD & 20 & (13.1, 4.96) & (22.05, 2.89) & (11, 9) & (13, 7) \\ \bottomrule
\end{tabular}
\end{table}

\subsection{Participant recruitment}
We recruited ninety-nine undergraduate and graduate participants from our institute for the study via email. Initially, an email with general information about the study was sent to all the institute students. The interested participants filled out the Google form, providing their demographic details and availability for the study. The inclusion criteria were that the participant should be over 18 years old and well-versed in English. Table \ref{tab:participant_characterstics} shows the demographic characteristics of the participants included in the analysis. As discussed later, forty-eight participants out of ninety-nine were dropped during the data analysis due to noisy data.

\subsection{Study procedure}
\begin{figure}[!h]
    \centering
    \begin{subfigure}{0.35\textwidth}
        \includegraphics [width=\textwidth]{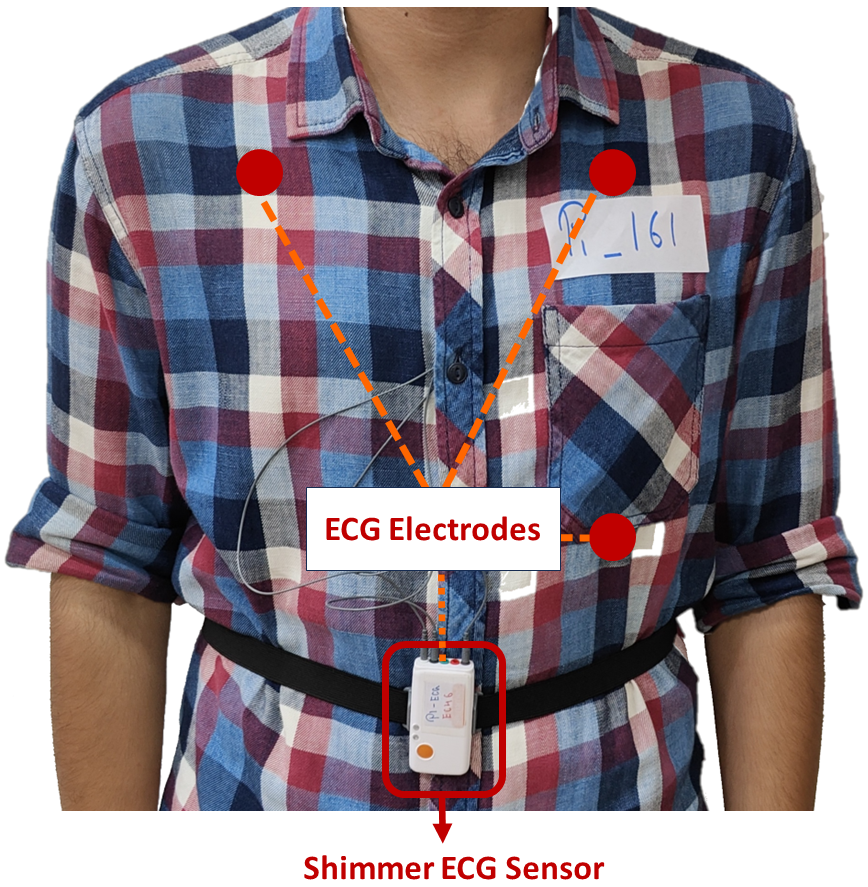}
        \caption{ECG sensor placement}
        \label{fig:sensor_placement}
    \end{subfigure} 
    \begin{subfigure}{0.55\textwidth}
        \includegraphics [width=\textwidth]{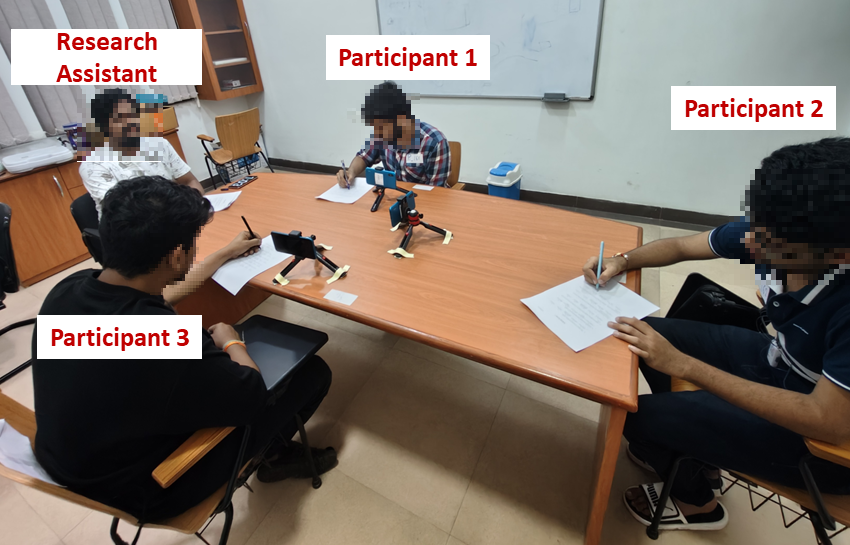}
        \caption{Seating arrangement during a study session.}
        \label{fig:sitting}
    \end{subfigure}
    \caption{Participants participating in the controlled lab study}
\end{figure}

The interested participants were grouped into trios by the research assistant (RA) while ensuring the group participants are from different academic batches and hence do not know each other prior to study. On the study day, three participants (P1, P2, and P3) visited one of the institute's rooms dedicated for the study. On arrival, the RA greeted and informed participants about the study verbally and using a participant information sheet (PIS), a form containing information such as study aim, time required, benefits, risks and confidentiality, etc. After reading PIS, the participants signed the informed consent form and filled out the Social Phobia Inventory (SPIN), a self-reported anxiety questionnaire (discussed in section~\ref{Self_reported_questionnaire}). Next, the RA requested participants to keep their smartphones on silent mode and placed a shimmer ECG sensor on their chest as shown in Figure~\ref{fig:sensor_placement}. Shimmer ECG sensors are designed for clinical trials and collect ECG data at a sampling rate of 1024. By default, the shimmer ECG sensor collects ECG data once placed on the chest. The participant's seating arrangements during a study session is shown in Figure~\ref{fig:sitting}.

\begin{figure}[!h]
  \centering
  \includegraphics[width=\textwidth, height=0.2\textheight, keepaspectratio]{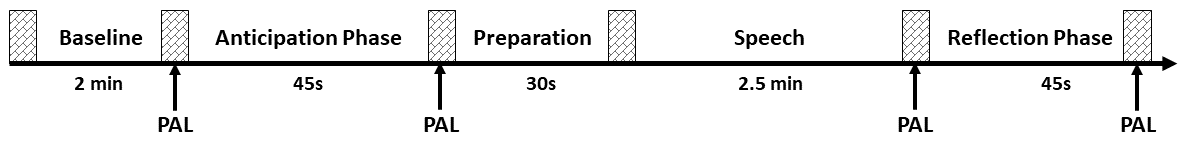}
  \caption{ 
 Sequence of different phases in the study. PAL denotes instances at which participants reported perceived anxiety levels (PAL) through surveys. Shaded blocks \includegraphics[width=0.20cm, height=0.35cm]{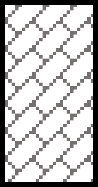}, represent instances whenever RA gave instructions to the participants.
  }
  \label{fig:study flow}
\end{figure}


Each study session comprised of a baseline and three other distinct phases, i.e., anticipation, speech activity, and reflection as shown in Figure~\ref{fig:study flow}. During the study, the RA noted each phase's start and end times. Following is a detailed description of each of these phases.
\begin{enumerate}
  \item \textit{\textbf{Baseline phase}}: During this phase, all three participants remained idle and refrained from engaging in any activity for 2 minutes. After 2 minutes, participants reported their perceived anxiety level (PAL) through a survey form. 
  
  \item \textit{\textbf{Anticipation Phase}}: During this phase, participants anticipated about the upcoming speech activity for 45 seconds. Specifically, the RA instructions were \textit{``Let's do the Speech activity. In this activity, each of you needs to speak for at least 2.5 minutes on a given topic. You will get 30 seconds to prepare the speech. Now, I want you to think or imagine for 45 seconds about this speech activity''}.
 A few participants asked the RA what exactly RA meant by anticipation, so the RA told them to think about potential speech topics and preparation strategy. Following the anticipation phase of 45 seconds, the RA requested participants to report their PAL by filling out a survey form again.
 
  \item \textit{\textbf{Activity Phase}}: During the activity phase, each participant gave a speech in front of the other two participants and RA for 2.5 minutes on an assigned topic. We chose speech as our anxiety-provoking activity while following the Trier Social Stress Test (TSST) guidelines. Speech is a well-known TSST activity like email-usage pattern~\cite{akbar2019email} or math test~\cite{paredes2018fast} and has been used to investigate the changes in the HR/HRV parameters~\cite{garcia2017autonomic,harrewijn2018heart,mauss2003autonomic,mauss2004there}. The RA instructions to P1 were, \textit{``Your speech topic is $<$assigned topic$>$. Take 30 seconds to prepare the speech''}. After 30 seconds, the RA asks P1 {\it ``please start your Speech''}. Following the speech activity,  P1 reported her PAL by completing a survey. Similarly, P2 and P3 complete the speech activity and survey forms one after the other. The speech topics were, ``Importance of Value Education'', ``Animal Zoos should be banned'', ``Effect of fake news on society'', etc.

  \item \textit{\textbf{Reflection Phase}}: During the reflection phase, participants introspect for 45 seconds about the recent speech activity, thinking about how it went. Specifically, RA instructions were,  \textit{``Now, could all of you think about the recent activity and how did it go in your mind?''} Following the reflection phase of 45 seconds, the participants again reported PAL by filling out a survey form. 
\end{enumerate}

After the Reflection phase, the RA thanked the participants for their participation and removed the shimmer sensors. The participants were provided a participation certificate and snacks after the study session.
Furthermore, the RA extracted ECG data from the shimmer sensors and stored it on the lab computer.

During the study, only one participant refused to continue with the study after reading the PIS. For each study session, we had two dummy participants as a backup to handle such situations where either a participant refuses to take part or does not appear on time. Dummy participants participated in seven study sessions to ensure study homogeneity. The dummy participant's data was later excluded during the data analysis. It took us four months to collect the data from a total of 99 participants. With this paper, we release our collected dataset\footnote{\url{https://osf.io/phze7/}} publicly.


\subsection{Ground Truth} \label{Self_reported_questionnaire}

We used two self-reported questionnaires during the study. The first is the standard SPIN questionnaire, and the second is our designed questionnaire. 

\subsubsection{Social Phobia Inventory (SPIN)}

\begin{wrapfigure}{r}{0.55\textwidth} 
\centering
  \includegraphics[scale=0.2]{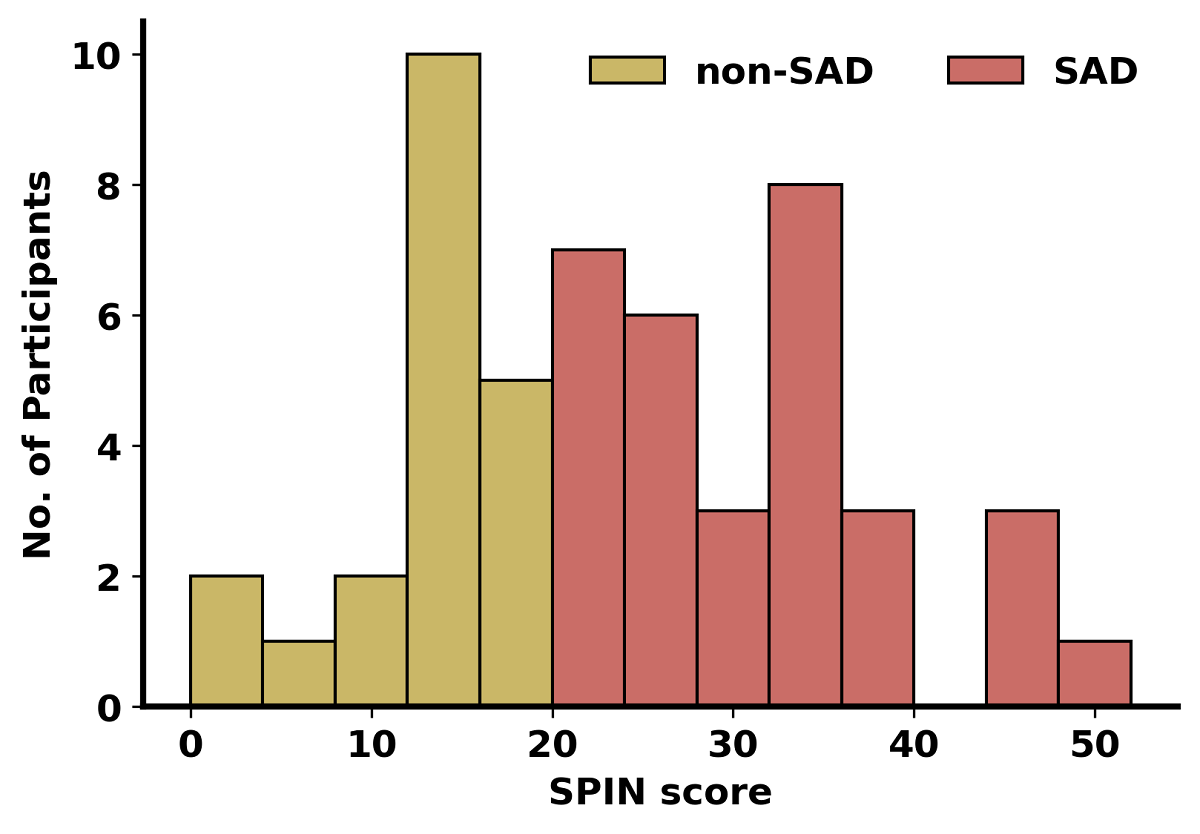}
  \caption{Distribution of SPIN Scores. Participants with SPIN scores greater than 20 were labeled as SAD, while those with scores of 20 or below were labeled as non-SAD.
  }
\label{fig: spin_distribution}
\end{wrapfigure}

Social anxiety can manifest in various forms (such as physical, psychological, and behavioral) across individuals in social situations \cite{stein2008social,ollendick2002developmental}. We used the SPIN questionnaire to assess the severity of SAD in participants. It is effective in screening and measuring the severity of SAD and is widely used for participant screening worldwide~\cite{connor2000psychometric,miranda2014anxiety, chukwujekwu2018validation}. Developed by the Department of Psychiatry and Behavioral Sciences at Duke University, SPIN consists of 17 items addressing issues such as ``fear of physical symptoms'', ``fear of negative evaluation'', and ``fear of uncertainty in social situations''. Participants were requested to respond to all 17 questions on a Likert scale from 0 to 4, where 0 indicated ``Not at all'' and 4 indicated ``Extremely''. The SPIN responses were collected before the placement of sensors, and participants were unaware of the activities they would later perform. Further, we used the SPIN responses during data analysis to understand the severity of the SAD of the participants and classify them into SAD and non-SAD groups. Figure \ref{fig: spin_distribution} shows the distribution of SPIN scores at a bin width of 4.

\subsubsection{Perceived Anxiety Level (PAL)}
\begin{figure}[!h]
    \begin{subfigure}{0.24\textwidth}
        \includegraphics [width=\textwidth]{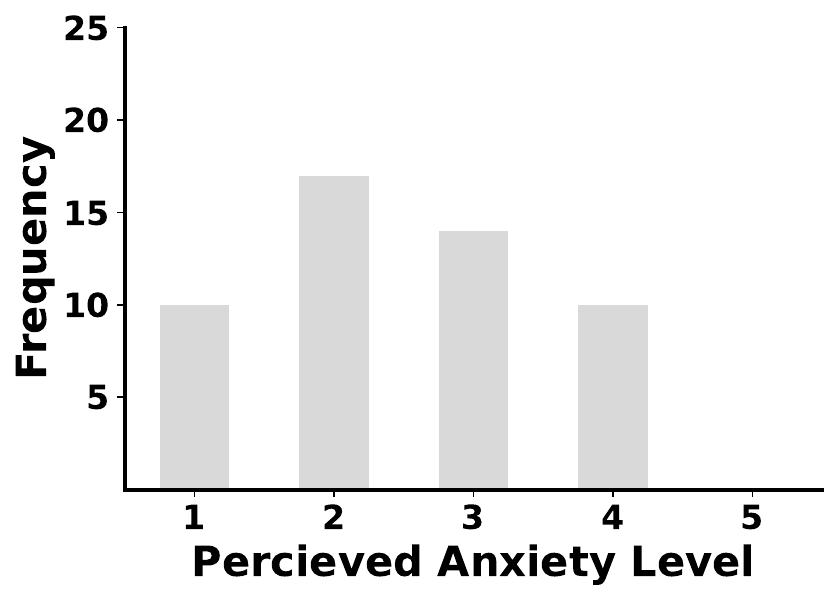}
        \caption{Baseline}
        \label{fig: hist_bp}
    \end{subfigure} 
        \begin{subfigure}{0.24\textwidth}
        \includegraphics [width=\textwidth]{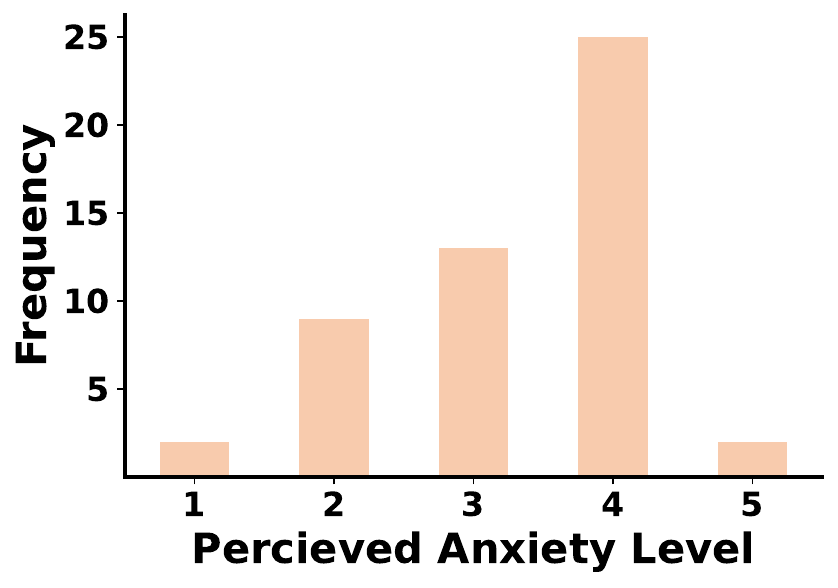}
        \caption{Anticipation}
        \label{fig: hist_atp}
    \end{subfigure} 
    \begin{subfigure}{0.24\textwidth}
        \includegraphics [width=\textwidth]{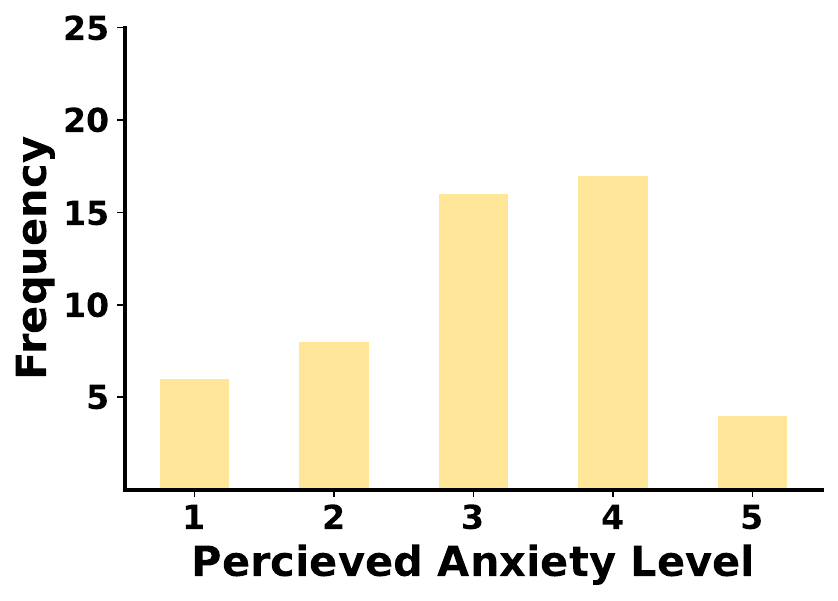}
        \caption{Speech activity}
        \label{fig: hist_ap}
    \end{subfigure} 
    \begin{subfigure}{0.24\textwidth}
        \includegraphics [width=\textwidth]{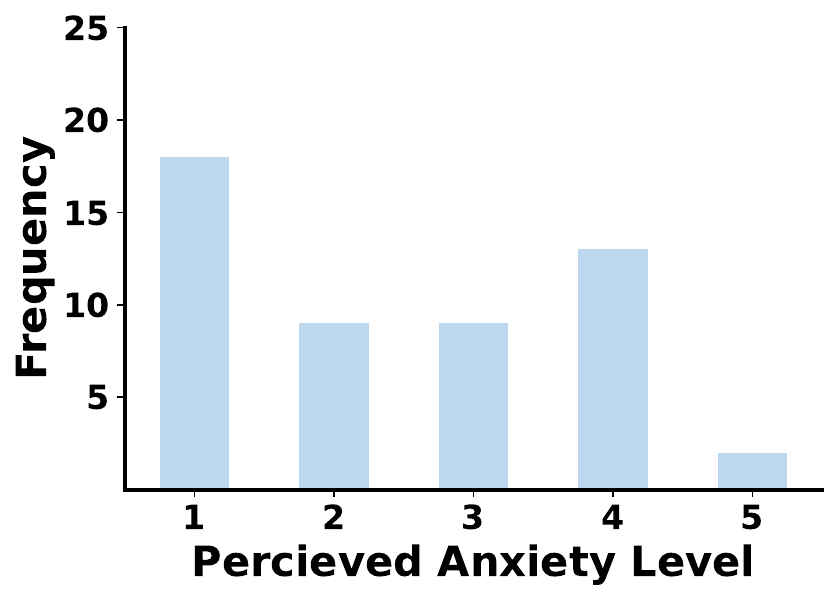}
        \caption{Reflection}
        \label{fig: hist_rp}
    \end{subfigure} 
    \caption{Distribution of perceived anxiety level (PAL) at the baseline and different phases (anticipation, speech activity, and reflection).}
    \label{fig:hist_sr}
\end{figure}

To assess participant's anxiety levels during the baseline and each phase (anticipation, activity, and reflection), we collected responses to a single self-reported question related to PAL as shown in Figure~\ref{fig: study_framework}. The rationale for using this single question was to understand how participants perceive different phases of an anxiety-provoking activity and how their anxiety levels change from one phase to another. Participants rated their PAL on a 5-point scale, where 1 indicates ``no anxiety'' and 5 indicates ``very high anxiety''. Figure \ref{fig:hist_sr} shows the distribution of participants' PAL at the baseline and different phases. Primarily, participants perceived highest anxiety in the anticipation phase.

\subsection{Study rationale} 

The study was designed by a collaborative team of psychiatrists and computer science professionals and was approved by the Institute's Review Board. It was carefully crafted to ensure the uniqueness of each phase, thereby enabling us to accurately assess the impact of speech activity on HR/HRV during three different phases compared to the baseline. The multiple-phase study design is innovative because it enables us to investigate whether physiological parameters, specifically  HR and  HRV, exhibit variations among participants during distinct phases: anticipation, activity, and reflection. Moreover, this design allows us to investigate whether HR/HRV parameters respond similarly among anxious and non-anxious participants during the speech activity. The anxiety triggers associated with each phase - ``not knowing the speech topic'', ``speaking in front of others'', and ``reflecting on the speech'' during anticipation, activity, and reflection, respectively - differ significantly from each other.

\begin{figure}[!h]
    \begin{subfigure}{0.33\textwidth}
        \includegraphics [width=\textwidth]{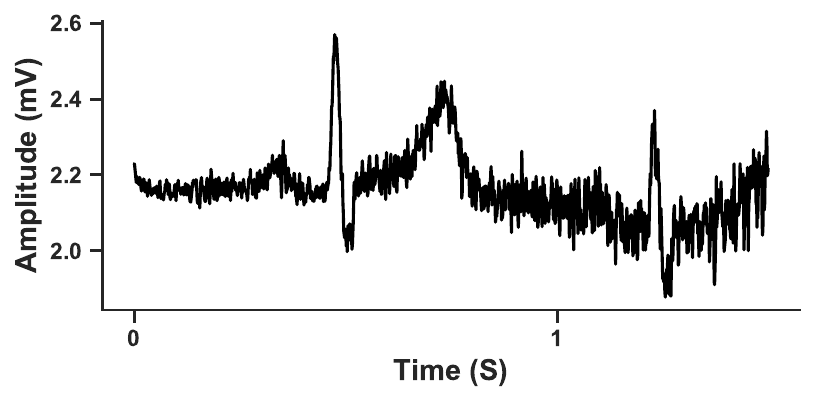}
        \caption{Raw ECG Signal}
        \label{fig: raw_ecg}
    \end{subfigure} 
        \begin{subfigure}{0.33\textwidth}
        \includegraphics [width=\textwidth]{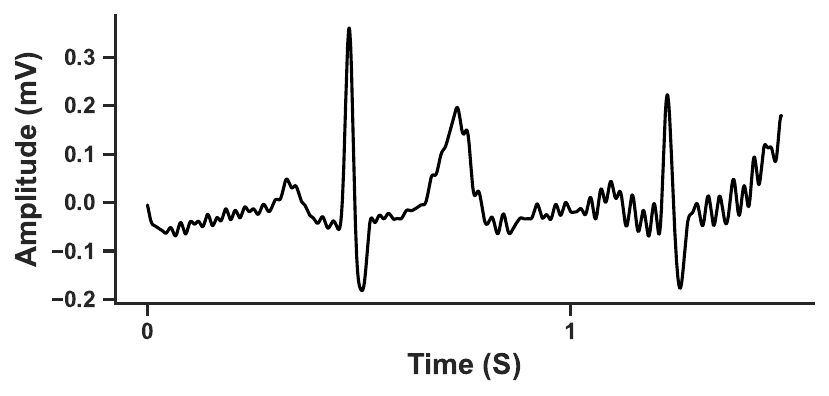}
        \caption{Filtered ECG Signal}
        \label{fig: filtered_ecg}
    \end{subfigure} 
    \begin{subfigure}{0.33\textwidth}
        \includegraphics [width=\textwidth]{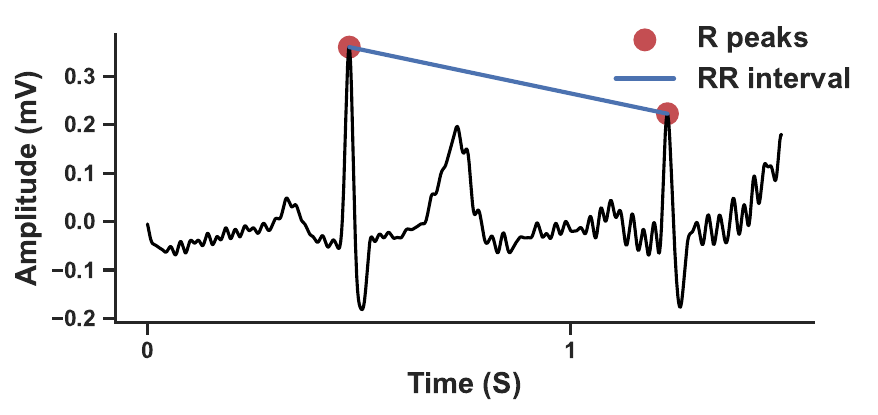}
        \caption{Filtered ECG Signal with Rpeaks}
        \label{fig: filtered_ecg_with_rpeaks}
    \end{subfigure} 
    \caption{Figures showing Raw ECG signal, filtered ECG signal, and filtered signal with R peaks.}
    \label{fig:signal_filtering}
\end{figure}

 \begin{figure} [!h]
    \centering
    \begin{subfigure}{0.45\textwidth}
        \includegraphics [width=\textwidth]{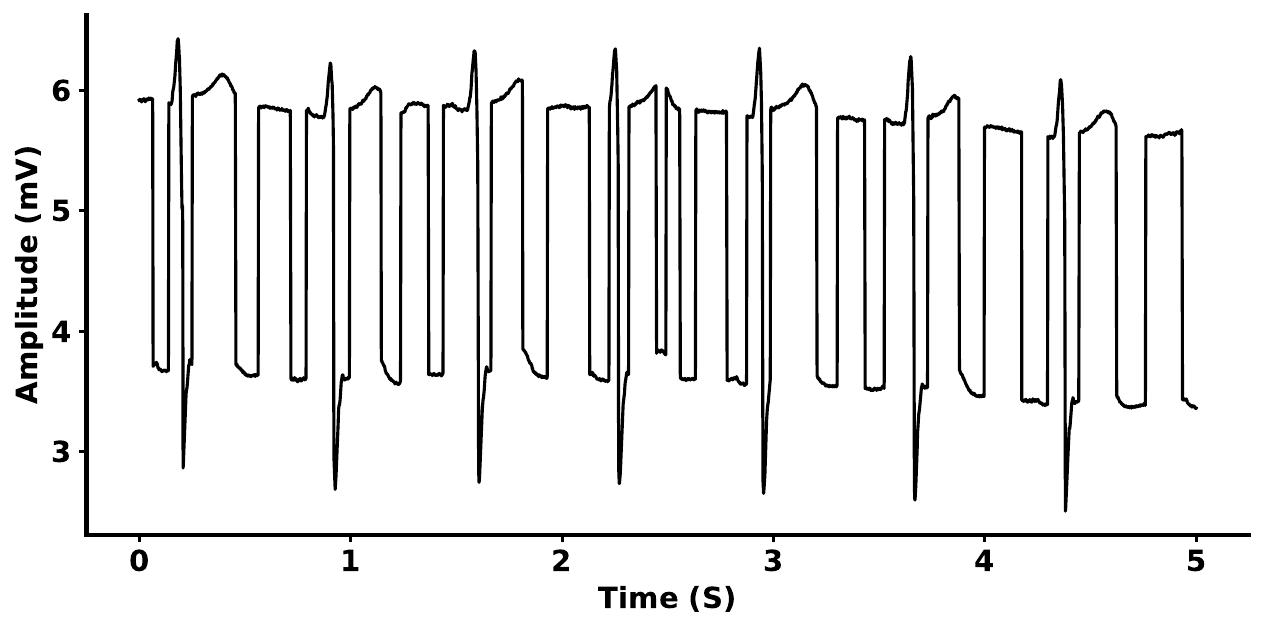}
        \caption{Signature of ECG signal due to faulty device}
        \label{fig:faulty_ecg}
    \end{subfigure} 
    \begin{subfigure}{0.45\textwidth}
        \includegraphics [width=\textwidth]{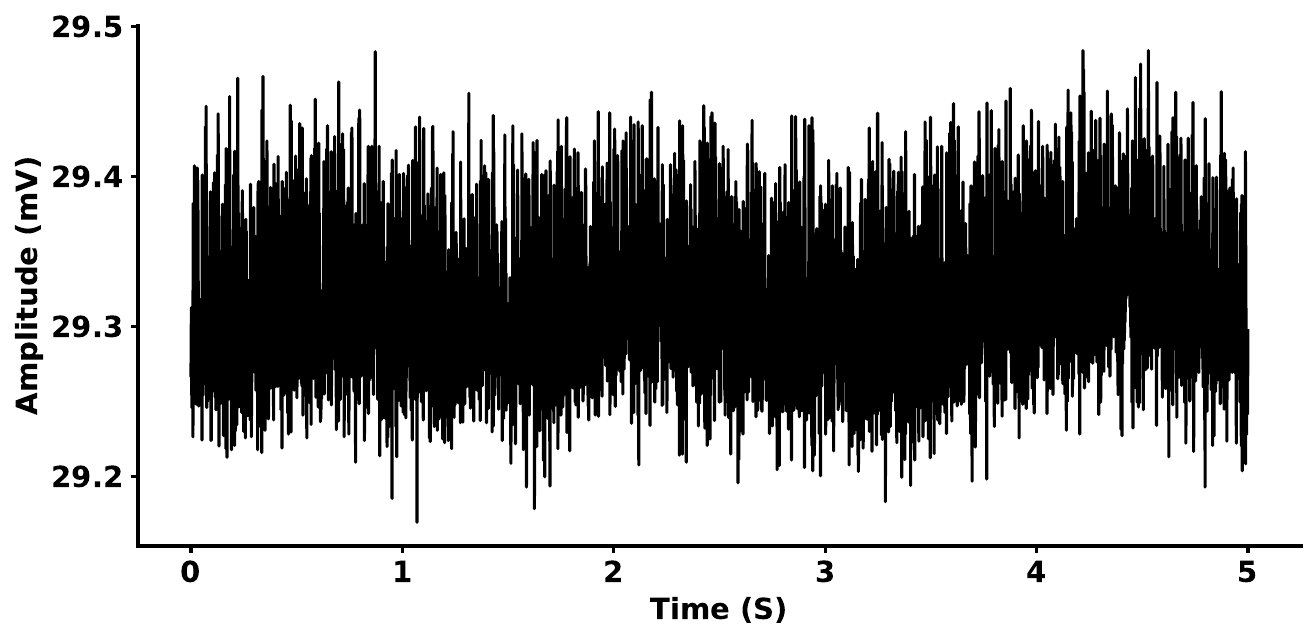}
        \caption{Signature of noisy ECG Signal}
        \label{fig:noisy_ecg}
    \end{subfigure}
    \caption{Examples of ECG signatures dropped during signal filtering from the study}
\end{figure}
\section{Analysis} \label{sec: Analysis}

\subsection{Data cleaning}
Before analyzing ECG data, filtering out noise and artifacts caused by equipment issues, body movements, or poor electrode contact is crucial. To achieve this, we applied a Finite Impulse Response (FIR) filter with a bandpass frequency range of 1-49 Hz to the recorded ECG signals. We compared the filtered signals to the ideal ECG waveform to evaluate the filter's effectiveness and found consistent results across all participants' data. Additionally, we utilized the Hamilton Segmenter algorithm for R peak detection \cite{hamilton2002open}. Figure \ref{fig:signal_filtering} illustrates the raw, filtered, and R peak detection plots for one participant's ECG data.

While filtering, we dropped 48 participants' data for further analysis due to various reasons, such as
({\it i}) The data was corrupted, as shown in Figure~\ref{fig:faulty_ecg}. This usually results due to loose connections or faulty skin electrodes. Two of the 48 participants were dropped due to faulty ECG data.
({\it ii}) The data was of poor quality even after filtering, as shown in Figure~\ref{fig:noisy_ecg}. Consequently, the R-peak detection algorithm either missed correct peaks or incorrectly labeled other points as R-peaks. Due to the noisy data, 46 participants were excluded from the analysis. This high exclusion rate was because noise was detected in at least one of the phases in the participant's ECG data. The phase-wise distribution of excluded participants' data is as follows: baseline (22), anticipation (14), activity (36), and reflection (17).

It is important to understand that ``noisy data'' refers to
 (i) instances where the entire phase's data was noisy, and 
 (ii) instances where there were noises interspersed within otherwise good signals.
 Both instances occurred in our data, and we chose to exclude them from further analysis for two primary reasons. Firstly, peak detection algorithms do not perform well when an entire phase's data is noisy even after filtering and yields incorrect HRV parameters. Secondly, dealing with noises interspersed within signals proved challenging because attempts to crop out the noisy portions are subjective and can result in biased results. These adjustments often altered R peak differences, impacting the overall results. So, the final dataset used in the analysis contained 51 participants data.

\begin{table}[!t]
\centering
\small
\caption{List of HRV and HR parameters used in this study.}
\label{tab:feature}
\begin{tabular}{@{}ll@{}}
\toprule
\textbf{Feature} & \multicolumn{1}{c}{\textbf{Explanation}} \\ \midrule
MeanNN & The average of RR intervals. \\
SDNN & The standard deviation of RR intervals. \\
RMSSD & The square root of the mean of squared successive differences between adjacent RR intervals. \\
MedianNN & The median of the RR intervals. \\
Prc20NN & The 20th percentile of the RR intervals. \\
pNN20 & The proportion of RR intervals greater than 20ms out of the total number of RR intervals. \\
HTI & The total number of RR intervals divided by the height of the RR intervals histogram. \\
TINN & An approximation of the RR interval distribution. \\
HF & The spectral power of high frequencies (0.15 to 0.4 Hz). \\
HFn & The normalized high frequency, obtained by dividing the low-frequency power by the total power. \\
LnHF & The log-transformed HF. \\
S	& The area of an ellipse described by SD1 and SD2, proportional to SD1SD2. \\
SD1	& The standard deviation perpendicular to the identity line, reflecting short-term RR interval fluctuations. \\
SD2	& The standard deviation along the identity line, representing long-term HRV changes. \\
SD1SD2	& The ratio of SD1 to SD2, indicating the balance between short-term and long-term HRV variations. \\
DFA$\alpha$1	& The monofractal detrended fluctuation analysis of the HR signals.\\
ApEn 	& The level of irregularity or randomness in the heartbeat intervals. \\
HR	& The number of times the heart beats per minute. \\ \bottomrule
\end{tabular}
\end{table}

\subsection{Feature extraction}
We extensively explored \cite{wang2023detecting,urrestilla2020measuring,shaffer2017overview,reyero2022heart} to extract various HRV features from the R peaks. Additionally, we incorporated all the features utilized by other relevant studies  \cite{wang2023detecting, alvares2013reduced, gaebler2013heart,tamura2013salivary,pittig2013heart,harrewijn2018heart,garcia2017autonomic,tolin2021psychophysiological,madison2021social,bailey2019moderating}. The motivation behind this decision was to align our research methodology with prior investigations, given the absence of established guidelines regarding anxiety-specific HRV features. The finalized set of features chosen for our analysis is presented in Table \ref{tab:feature}, along with their respective explanations. These features include time domain computed metrics such as MeanNN, SDNN, RMSSD, MedianNN, Prc20NN, pNN20, HTI, TINN, and HR, as well as frequency domain computed features like HF, HFn, and LnHF. Additionally, the table includes nonlinear indices such as S, SD1, SD2, SD1/SD2, DFA$\alpha$1, and ApEn. We did not include any low-frequency HRV parameters in our study, as their validity has been questioned in previous studies \cite{tamura2013salivary, cheng2022heart}.

To extract HRV features (i.e., time, frequency, and nonlinear indices of HRV), we used the Neurokit Python package \cite{makowski2021neurokit2}. We fed the R-peak indices, obtained using the Hamilton Segmenter on a cleaned ECG signal, into the respective functions for time, frequency, and nonlinear indices separately. These functions provided aggregated time domain, frequency domain, and nonlinear indices for each ECG signal. This process was repeated for each participant across all phases (i.e., baseline, anticipation, activity, and reflection). As a result, we obtained the time domain, frequency domain, and nonlinear indices for each phase.

\subsection{Participant's grouping} 

Following existing studies~\cite{connor2000psychometric, miranda2014anxiety, chukwujekwu2018validation, lauria2016primary}, we categorized 51 participants into SAD and non-SAD groups, using a threshold of 20 on the SPIN score. Participants with scores exceeding 20 were labeled as SAD, while those with scores below were classified as non-SAD. Consequently, the total participant pool of 51 participants was divided into two distinct groups: 31 SAD and 20 non-SAD participants. Based on SPIN scores and at a significance level of 0.05, the computed effect size difference was found to be 2.68. This substantial effect size indicates a significant distinction between the two groups.

 \begin{wrapfigure}{r}{0.5\textwidth} 
\centering
  \includegraphics[scale=0.45]{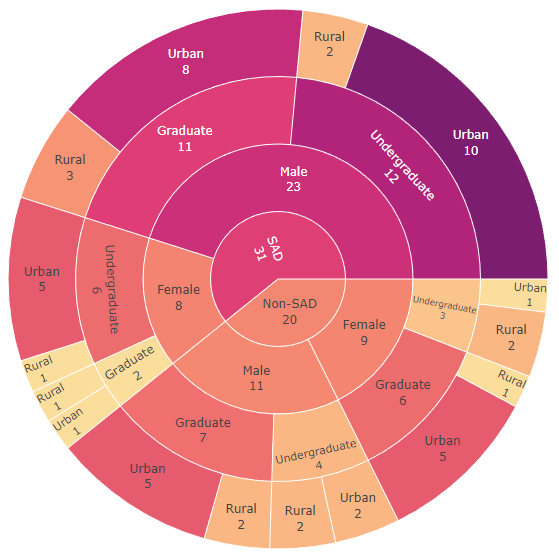}
  \caption{Distribution of 51 participants.}
\label{fig: pc_distribution}
 \end{wrapfigure}
 
The sunburst chart in Figure~\ref{fig: pc_distribution} represents the participant distribution across different levels. There were 31 SAD and 20 non-SAD participants. Among the SAD group, there were 28 males and eight females, while the non-SAD group had 11 males and nine females. Additionally, the distribution of participants' home residence within the SAD group was as follows: 24 participants resided in urban areas and seven in rural areas, while in the non-SAD group, 19 participants were from urban areas and seven from rural areas.

\subsection{Correlation analysis} \label{subsection: correlation}

We performed correlation analysis to explore the relationships between the aggregate HR/HRV features and the self-reported measures (i.e., pre-baseline SPIN and PAL) at different study phases. This analysis aimed to provide insights into the direction and strength of the associations between HR/HRV and the self-reported measures. 
Now, we will discuss the results obtained through correlation analysis in two aspects: (i) the correlations between pre-baseline SPIN scores and HR/HRV collected during baseline and different phases, (ii) the correlations between PAL and HR/HRV collected during baseline and various phases of the study, and (iii) the correlations between pre-baseline SPIN and PAL during baseline and various phases of the study. Following are the findings.

\begin{figure}[!h]
\centering
  \includegraphics[scale=0.22]{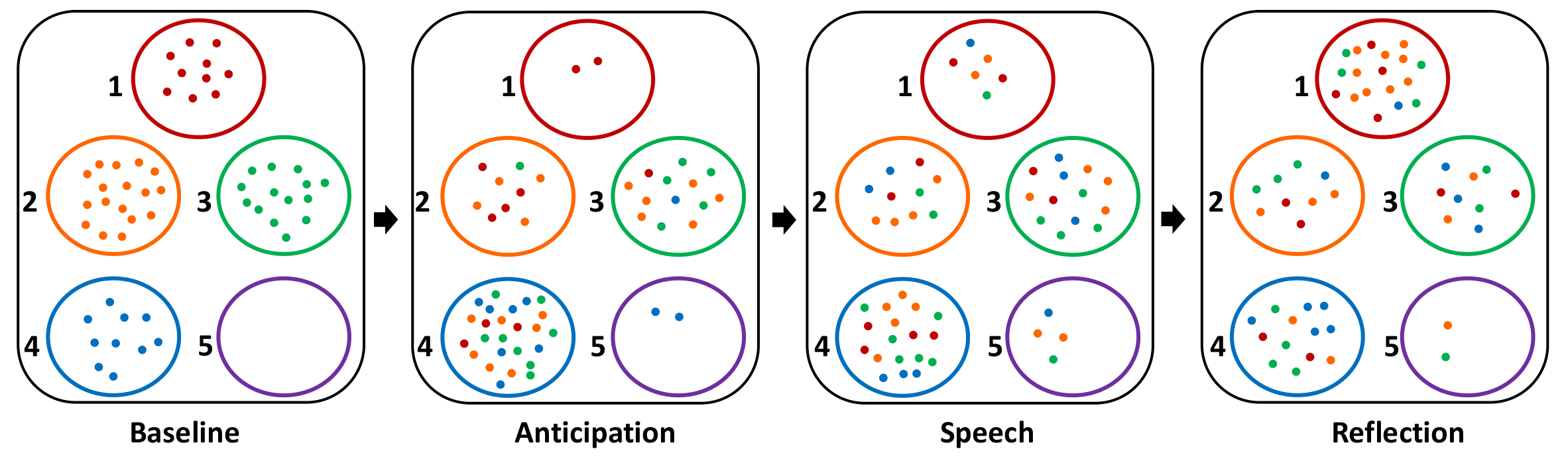}
  \caption{
Each circle represents a cluster, and each dot represents a participant. Red, orange, green, blue, and purple dot colors at ``Baseline'' denote the perceived anxiety levels of 1, 2, 3, 4, and 5, respectively. 
The figure highlights the transition of participants' anxiety levels from the Baseline to a specific phase (i.e., Anticipation, Speech, and Reflection). [{\bf Best viewed in color}]    } 
\label{fig:self_report_visual}
\end{figure}

\begin{itemize}
    \item \textit{Correlation results of HR/HRV features of each phase with SPIN} - The correlation analysis revealed that most HRV features exhibited negative correlations with the SPIN score, regardless of the phase. Specifically, MeanNN, MedianNN, Prc20NN, HF, HFn, and LnHF displayed correlations between -0.1  and -0.40, while other features fell between -0.09 and +0.09. Heart rate (HR) showed positive correlations with SPIN across all phases, with correlations ranging from +0.19 to +0.20. These findings suggest that participants with higher SPIN scores tend to have lower HRV and higher HR, highlighting a consistent relationship between social anxiety levels and these physiological measures syncing with the literature that SAD participants exhibit lower HRV and higher HR \cite{alvares2013reduced}.
    
    \item \textit{Correlation results of HRV and PAL of each phase} - Nearly all HRV measures exhibited negative correlations with their respective PAL in each phase. Specifically, HF, HFn, and LnHF displayed correlations within the range of -0.1 to -0.22, while other HRV features fell within the range of -0.09 to +0.15. We observed a high positive correlation (0.22) between HR and PAL during the speech activity, whereas, for other phases, these correlations were less pronounced (baseline ($\rho =0.04$), anticipation ($\rho =0.09$), and reflection ($\rho =0.05$)). These results align with the trends observed in the SPIN scores, except for the relationship between HR and anxiety, which exhibited some differences.
    
    \textbf{\textit{Note:}} Building on the above two analyses, it is worth highlighting that all high-frequency HRV features, i.e., HF, HFn, and LnHF, displayed consistent negative correlations in the baseline, anticipation, and reflection phases. However, these measures exhibited a positive correlation during the speech activity.

\begin{figure}[!h]
    \centering
    \begin{subfigure}{0.235\textwidth}
        \includegraphics [width=\textwidth]{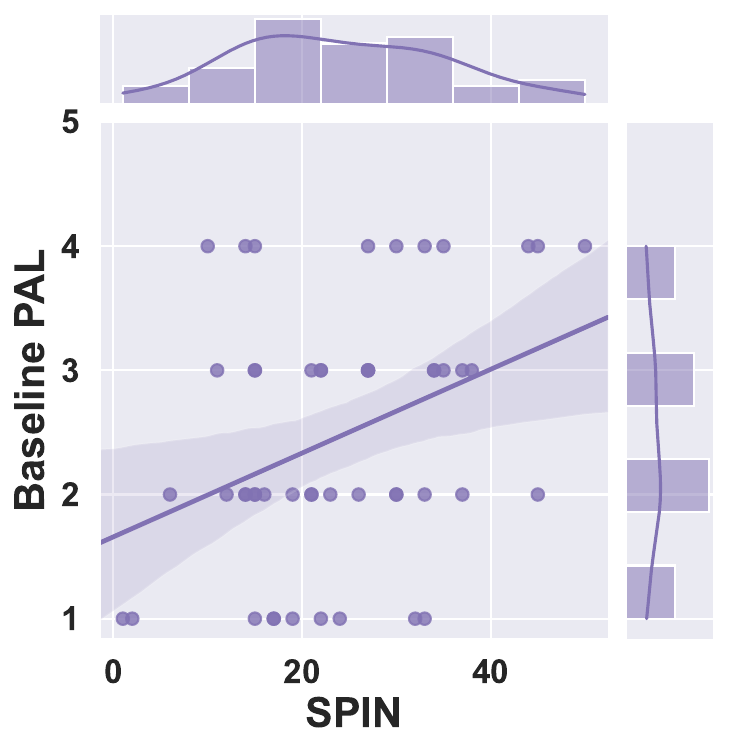}
        \caption{Baseline}
        \label{fig:Joint_plot_SPin_Baseline_SR}
    \end{subfigure} 
        \begin{subfigure}{0.235\textwidth}
        \includegraphics [width=\textwidth]{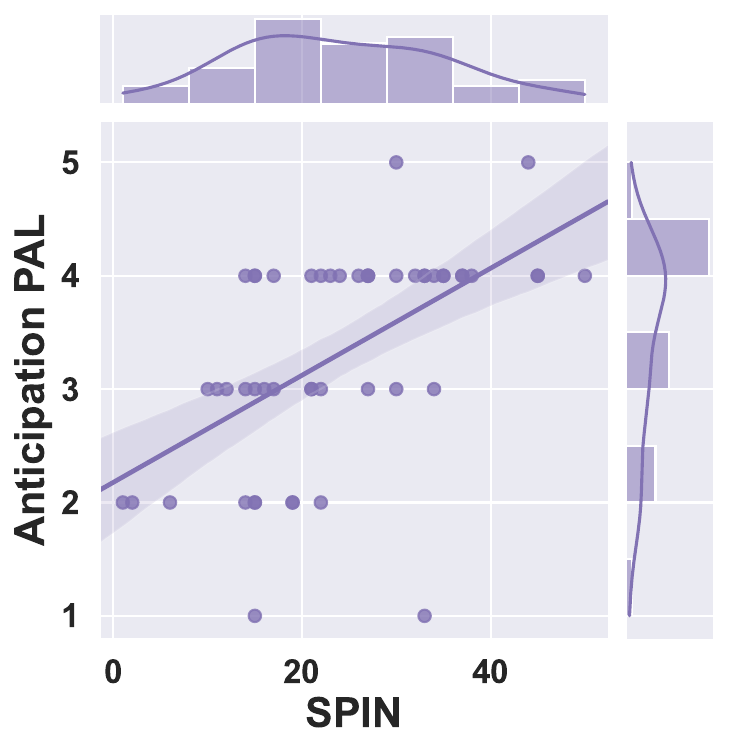}
        \caption{Anticipation Phase}
        \label{}
    \end{subfigure} 
    \medskip 
    \begin{subfigure}{0.235\textwidth}
        \includegraphics [width=\textwidth]{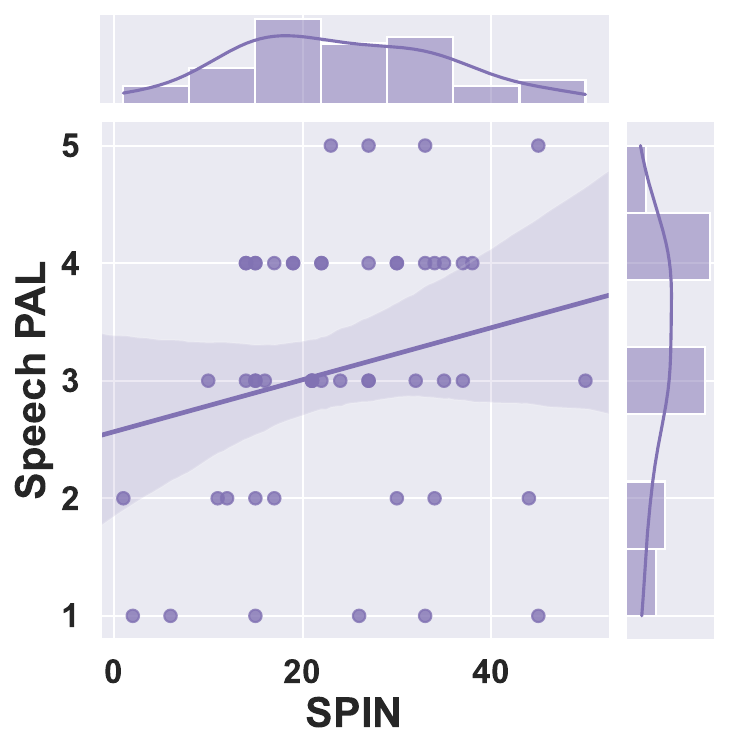}
        \caption{Speech}
        \label{}
    \end{subfigure} 
    \begin{subfigure}{0.235\textwidth}
        \includegraphics [width=\textwidth]{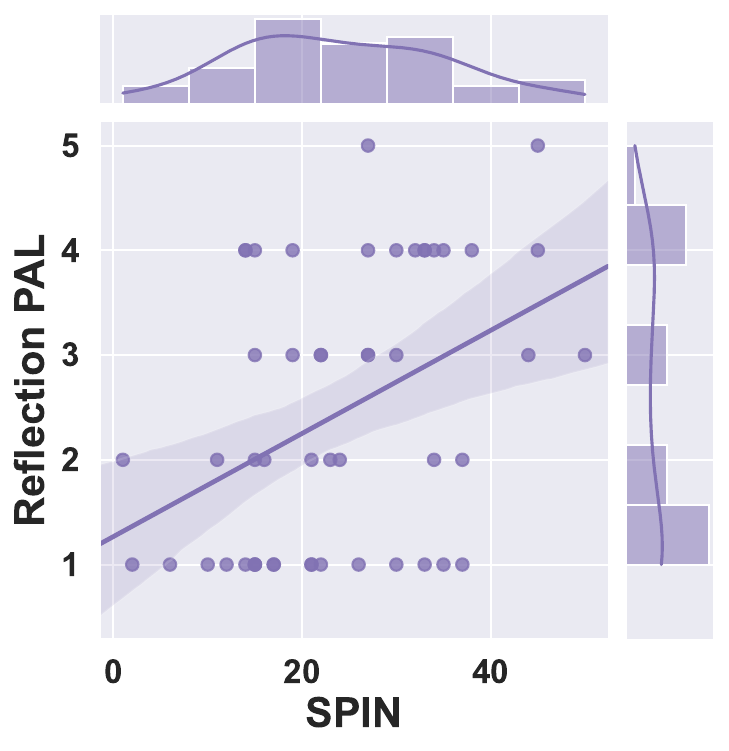}
        \caption{Reflection Phase}
        \label{}
    \end{subfigure}
    \caption{Relationship between SPIN and PAL in different phases (Baseline, Anticipation, Speech, and Reflection). The X-axis represents the SPIN score, and the Y-axis represents the PAL.}
    \label{fig: joint_plot_spin_sr]}
\end{figure}

\item \textit{Correlation results of SPIN score with PAL of each phase} - The correlation between the SPIN and PAL  exhibited variation across different phases. The highest correlation ($\rho = 0.56$) was observed during the anticipation phase, whereas the lowest was noted during the speech ($\rho = 0.22$). This finding suggests that participants with low SPIN scores would have perceived higher anxiety during the speech activity.  

Figure~\ref{fig: joint_plot_spin_sr]} showing four joint plots illustrates the relationship between SPIN and PAL at different phases. For example, in Figure \ref{fig:Joint_plot_SPin_Baseline_SR}, the X and Y axes represent SPIN scores and PAL at the baseline. Each dot corresponds to participants' SPIN and PAL, and the fitted regression line summarizes the relationship between the dots. Additionally, the joint plots display the underlying distribution of SPIN and PAL in the top and ride side panels. Top distributions in all the joint plots are the same, as the SPIN was recorded once during the study. However, right-side distributions keep changing in different phases as PAL was recorded separately in each phase. The distributions show that during the baseline, anticipation, speech, and reflection phases, most participants reported anxiety levels of 2, 4, 4, and 1, respectively. \textit{This shows that anticipation and speech phases increased anxiety in the participants}.   

 Furthermore, we use Figure \ref{fig:self_report_visual} to understand the transitions in participants PAL from baseline to other phases (i.e., Anticipation, Speech, and Reflection). Circles labeled 1, 2, 3, 4, and 5 represent clusters with anxiety levels of 1, 2, 3, 4, and 5, respectively. Dots within the clusters represent participants reporting a particular anxiety level. For example, the figure shows that 17 participants reported an anxiety level of 2 in the baseline phase. However, out of the 17 participants, (i) during anticipation,  six moved to cluster 3, and seven moved to cluster 4; (ii) during speech,  two moved to cluster 1, four moved to cluster 3, five moved to cluster 4, and two moved to cluster 5; (iii) during reflection,  nine moved to cluster 1, two moved to cluster 3, two moved to cluster 4, and one moved to cluster 5. 
  Overall, the transitions from baseline to other phases further confirm that the PAL increased during anticipation and speech compared to the baseline phase. These transitions also highlight that the anticipation increased the anxiety in the participants higher than the actual speech activity itself.

    \end{itemize}

\subsubsection{Ablation Study}

This ablation study aimed to determine whether the correlation between HRV and self-reports (SPIN and PAL) differs between males and females. The rationale behind this investigation was motivated by existing literature, which suggests that females typically have lower HRV and higher HR than males \cite{geovanini2020age}. To explore this, we first conducted a correlation analysis between SPIN and HRV for males, followed by a similar analysis for females. We repeated a similar correlation analysis between males' PAL and HRV followed by females.

Our findings revealed that both males and females exhibit a negative correlation with the self-reports (both SPIN and PAL).
However, the negative association between HRV and self-reports was stronger in females. For instance, the correlation between SPIN and SDNN in males was 0.05, -0.02, -0.04, and -0.09 for baseline, anticipation, activity, and reflection phases, respectively. In contrast, the corresponding correlations for females were -0.23, -0.37, -0.58, and -0.08. We observed a similar pattern on analyzing the correlation between PAL and SDNN. These results suggest that gender differences play a significant role in the relationship between HRV and anxiety, highlighting the importance of considering gender factors while modeling.

\subsection{Hypothesis testing}

 The collected ECG data with demographic information (i.e., gender and location) has a multilevel structure. So, to address our hypotheses, we have used a multilevel modeling approach.
Multilevel modeling, also known as hierarchical linear modeling or mixed-effects modeling, is a statistical model used to analyze data that follows a nested, clustered, or hierarchical structure \cite{leyland2001multilevel,bristol_multilevel,leyland2020multilevel}. It allows for estimating both within and between-group effects, capturing the complex relationships within the data. The predictor variables in such models can be modeled as fixed or random effect variables. A fixed effect variable represents the average relationship between the predictor and the response variable, assuming this relationship is consistent across all groups or levels in the data. In contrast, a random effect variable accounts for variability between groups or levels, capturing how the relationship between the predictor and the response variable might differ across these groups or levels. We performed multilevel statistical analyses using the `lme4.0' package of R language \cite{bates2014fitting}.


Tables~\ref{tab:h1}, \ref{tab:h2}, \ref{tab:h3} provide a comprehensive summary of the changes in HRV, HR, and self-reported data (SPIN and PAL), along with their respective significance values, as the outcomes of testing hypotheses 1, 2, and 3 respectively. The Table~\ref{tab:h1} presents a clear overview of the statistical significance of these changes, shedding light on the impact of different phases on HRV, HR, and PAL in the study. Additionally, Table~\ref{tab:h2} compares with existing literature, demonstrating how our results align with previous research in this field. Now, we will discuss the approach employed and present the results for each of the hypotheses.

\noindent\textit{\textbf{H1:}}
We tested our hypothesis 1 with the following model: 
\begin{verbatim}
 lmer(HRV ~ phase + (1|group) + (1|pid) + (1|program) + (1|gender) + (1|location), dataset)
\end{verbatim}
Where {\tt HRV} refers to different HRV features mentioned in the Table~\ref{tab:feature}, {\tt phase} has four levels (i.e., baseline, anticipation, speech activity, and Reflection),  {\tt group} has two levels (i.e., SAD, non-SAD), {\tt pid} refers to participant-id,  {\tt program}  has two levels (i.e., undergraduate, graduate), {\tt gender} has two levels (i.e., male, female),  {\tt location} has two levels (i.e., rural, urban). We ran the above model for each of HRV features and HR separately. 

In the above model, the {\tt phase} variable is fixed (with baseline as reference) while the remaining features are random effects. We treat it as a fixed effect because we are interested in the overall average effect of each phase on HRV across all participants, regardless of their group membership or demographic characteristics. This allowed us to isolate and analyze changes in HRV/HR while accounting for other features such as age, gender, location, and groups. As SPIN reporting can be biased, we considered {\tt group} as a random effect. Table~\ref{tab:h1} reports the direction (increase with $\uparrow$ and decrease with $\downarrow$) and the significance of reported results. 

The results show a significant decrease in HRV features (SDNN, HTI, TINN, HF, SD1, and ApEn) and significant increases in HRV features HF, HFn, and LnHF during the anticipation phase compared to the baseline. Similarly, we observed significant reductions in HRV features (MeanNN, RMSSD, MedianNN, Prc20NN, pNN20, TINN, LnHF, S, SD1, and SD1SD2) and a significant increase in HR during the speech phase compared to the baseline. Conversely, during the reflection phase, we found a significant increase in HRV features (MeanNN, MedianNN, Prc20NN, HF, HFn, LnHF) and a significant decrease in HRV features (HTI, TINN, and ApEn) compared to the Baseline.


In summary, our findings indicate that during the anticipation and speech phases, HRV decreases, and HR increases compared to the baseline. However, this pattern reverses during the reflection, suggesting that anticipation and speech activity induced anxiety in all participants. 

\begin{table} [!t]
\caption{Results of hypothesis 1, 2, and 3, where (i) Acronyms ATP and RP denote Anticipation and Reflection, respectively, (ii) Arrows, $\uparrow$ and $\downarrow$ represent increase and decrease, respectively, (iii) Significance levels `***', `**', `*',  `.' correspond to 0.001, 0.01, 0.05, and 0.1, respectively, (iv) $\leftrightarrow$ means  that direction (increase or decrease) not reported by authors, and ``-'' means no literature exists.}
\centering
\subcaptionbox{Hypothesis 1 results \label{tab:h1}}{
\begin{tabular}{llll}
\toprule
\textbf{HRV} & \textbf{ATP} & \textbf{Speech} & \textbf{RP} \\ \midrule
MeanNN & $\downarrow$ & $\downarrow$*** & $\uparrow$*  \\
MedianNN & $\downarrow$. & $\downarrow$*** & $\uparrow$ *  \\
SDNN & $\downarrow$* & $\downarrow$ & $\downarrow$ \\
RMSSD & $\downarrow$. & $\downarrow$** & $\downarrow$  \\
Prc20NN & $\downarrow$ & $\downarrow$*** & $\uparrow$* \\
pNN20 & $\uparrow$ & $\downarrow$*** & $\uparrow$ \\
HTI & $\downarrow$*** & $\uparrow$ & $\downarrow$*** \\
TINN &$\downarrow$*** & $\downarrow$. & $\downarrow$*** \\ 
HF & $\uparrow$**& $\downarrow$. & $\uparrow$*** \\
HFn & $\uparrow$*** & $\uparrow$ & $\uparrow$*** \\
LnHF & $\uparrow$*& $\downarrow$** & $\uparrow$**\\ 
S & $\uparrow$* & $\downarrow$* & $\downarrow$ \\
SD1 & $\downarrow$* & $\downarrow$** & $\uparrow$ \\
SD2 & $\downarrow$** & $\downarrow$ & $\downarrow$ \\
SD1SD2 & $\downarrow$ & $\downarrow$** & $\uparrow$ \\
ApEn & $\downarrow$*** & $\downarrow$ & $\downarrow$*** \\
DFA$\alpha$1 & $\uparrow$ & $\uparrow$ & $\uparrow$ \\
HR & $\uparrow$ & $\uparrow$*** & $\downarrow$. \\ \bottomrule
\end{tabular}
}
\hfill
\subcaptionbox{Hypothesis 2 results \label{tab:h2}}{
\begin{tabular}{lll}
\toprule
\textbf{HRV} & \textbf{Groups} & \textbf{Literature}  \\ \midrule
MeanNN & $\downarrow$ & - \\
MedianNN & $\downarrow$ & - \\
SDNN & $\downarrow$. & $\downarrow$ \cite{alvares2013reduced}\\
RMSSD & $\downarrow$* & $\downarrow$ \cite{alvares2013reduced,garcia2017autonomic, harrewijn2018heart, madison2021social}, $\leftrightarrow $ \cite{tolin2021psychophysiological,tamura2013salivary, bailey2019moderating} \\
Prc20NN & $\downarrow$ & - \\
pNN20 & $\uparrow$ & - \\
HTI & $\downarrow$ & - \\
TINN & $\downarrow$ & - \\ 
HF & $\downarrow$ & $\downarrow$ \cite{alvares2013reduced,pittig2013heart,gaebler2013heart,tolin2021psychophysiological, harrewijn2018heart} \\
HFn & $\downarrow$ & - \\
LnHF & $\downarrow$ & - \\ 
S & $\downarrow$* & - \\
SD1 & $\downarrow$* & - \\
SD2 & $\downarrow$ & - \\
SD1SD2 & $\downarrow$ & - \\
ApEn & $\downarrow$ & - \\
DFA$\alpha$1 & $\uparrow$ & $\uparrow$ \cite{alvares2013reduced} \\
HR & $\uparrow$ & $\uparrow$ \cite{miranda2014anxiety,mauss2003autonomic,garcia2017autonomic,tolin2021psychophysiological,bailey2019moderating, grossman2001gender}, $\leftrightarrow $ \cite{gaebler2013heart, tamura2013salivary, pittig2013heart} \\ \bottomrule
\end{tabular}
}

\vspace{0.7cm}
\subcaptionbox{Hypothesis 3 results \label{tab:h3}}{
\begin{tabular}{llllll}
\toprule
 &\textbf{Anticipation} & \textbf{Speech} & \textbf{Reflection} & \textbf{Groups} & \textbf{Literature}  \\ \midrule
 PAL & $\uparrow$*** & $\uparrow$*** & $\downarrow$ & $\uparrow$ ** & $\uparrow$ \cite{mauss2003autonomic,garcia2017autonomic,mauss2004there} \\ \bottomrule
\end{tabular}
}
\label{tab:hypothesis_table}
\end{table}

\noindent\textit{\textbf{H2:}}
We tested our hypothesis 2 with the following model: 
\begin{verbatim}
 lmer(HRV ~ group + phase + (1|pid) + (1|program) + (1|gender) + (1|location), dataset)
\end{verbatim}
In this model, we treated {\tt phase} and {\tt group} as fixed variables, while the remaining variables were considered random effects. The baseline in {\tt phase} and non-SAD in {\tt group} was set as the reference. We ran the above model for each HRV feature and HR separately. Our result in Table~\ref{tab:h2} shows that most HRV features decreased in the SAD participants compared to non-SAD participants irrespective of the phases. However, only RMSSD, S, and SD1 were found significant at a 0.05 significance level. Furthermore, we found an increase in HR in the SAD group; however, the result was not significant enough to be accepted. 



\noindent\textit{\textbf{H3:}} Hypothesis 3 focuses on PAL instead of the HRV/HR features. Specifically, it examines how PAL changes across different phases and whether SAD participants consistently report higher PAL. Here, we considered two sub-cases:

\textit{Case 1}: We treated the {\tt PAL} as the dependent variable, {\tt phase} as fixed (with baseline as reference), and the remaining features as random effects in the following model:
\begin{verbatim}
 lmer(PAL ~ phase + (1|group) + (1|pid) + (1|program) + (1|gender) + (1|location), dataset)
\end{verbatim}
Results in Table~\ref{tab:h3}  show that PAL increased significantly during the anticipation and speech phases compared to the baseline.
Additionally, a decrease in PAL was observed in the reflection phase, although the result was not statistically significant.  
  
  \textit{Case 2}: We treated the {\tt PAL} as the dependent variable, {\tt phase} (with baseline as reference) and {\tt group} (with non-SAD as reference) as fixed, and the remaining features as random effects in the following model:
  \begin{verbatim}
 lmer(PAL ~ group + phase + (1|pid) + (1|program) + (1|gender) + (1|location), dataset)
\end{verbatim}
  Table~\ref{tab:h3} results show that the perceived anxiety of SAD participants is significantly higher than that of non-SAD participants. 

\subsubsection{Ablation Study}

This ablation study determined the influence of gender on HRV features and PAL. Therefore, we analyzed changes in HRV and PAL by keeping the {\tt gender} (with male as reference) and {\tt phase} (with baseline as reference) as fixed variables and the remaining variables (i.e., group, participant id, program, and location) random in the following model:

  \begin{verbatim}
  lmer(HRV/PAL ~ phase + gender + (1|pid) + (1|program) + (1|groups) + (1|location), dataset)
\end{verbatim}

With male as the reference category, results in Table \ref{tab:female} show that females exhibit lower HRV, higher HR, and higher PAL than males irrespective of phases. Moreover, to gain insights into the differences between participants with and without social anxiety within genders, we treated HRV/PAL as a dependent variable and {\tt phase} (with baseline as reference), {\tt gender} (with male as reference), and {\tt group} (with non-SAD as reference) as fixed variables, and the remaining variables (participant id, program, and location) as random in the following model:

  \begin{verbatim}
  lmer(HRV/PAL ~ groups + phase + gender + (1|pid) + (1|program) + (1|location), dataset)
\end{verbatim}

Consistent with our previous results, we observed that females consistently displayed lower HRV than their male counterparts shown in Table \ref{tab:both_female_and_sad}. Additionally, our results show that including gender as a fixed variable increased the number of distinguishing HRV parameters (including SDNN, RMSSD, TINN, S, SD1, and SD2) between SAD and non-SAD groups. However, from hypothesis 2 testing, we found that only SDNN, RMSSD, S and SD1 could significantly differentiate the SAD and non-SAD groups.

\begin{table} [!h]
\caption{Changes in HRV, HR, and Self-reported anxiety analysis with gender as a fixed variable.
Where (i) Arrows, $\uparrow$ and $\downarrow$ represent increase and decrease, respectively, (ii) Significance levels `***', `**', `*',  `.' correspond to 0.001, 0.01, 0.05, and 0.1, respectively}
\centering
\subcaptionbox{Increase/decrease with gender as a fixed variable.~\label{tab:female}}{
\begin{tabular}{p{3cm} p{2cm}}
\toprule
\textbf{Features} & \textbf{Female} \\ \midrule
MeanNN & $\downarrow$* \\
MedianNN & $\downarrow$*  \\
SDNN & $\downarrow$*  \\
RMSSD & $\downarrow$*  \\
Prc20NN & $\downarrow$.  \\
pNN20 & $\downarrow$  \\
HTI & $\downarrow$*  \\
TINN &$\downarrow$  \\ 
HF & $\downarrow$  \\
HFn & $\uparrow$  \\
LnHF & $\downarrow$  \\ 
S & $\downarrow$*  \\
SD1 & $\downarrow$*  \\
SD2 & $\downarrow$*  \\
SD1SD2 & $\uparrow$  \\
ApEn & $\uparrow$  \\
DFA$\alpha$1 & $\uparrow$ \\
HR & $\uparrow$.  \\ 
Self-reported anxiety (i.e., PAL) &  $\uparrow$ \\ \bottomrule
\end{tabular}
}
\hspace{1 cm}
\subcaptionbox{Increase/decrease with both gender and groups as fixed variables. \label{tab:both_female_and_sad}}{
\begin{tabular}{p{3cm} p{1.5cm} p{1.5cm}}
\toprule
\textbf{Features} & \textbf{Female} & \textbf{SAD}  \\ \midrule
MeanNN & $\downarrow$* & $\downarrow$ \\
MedianNN & $\downarrow$* & $\downarrow$   \\
SDNN & $\downarrow$* & $\downarrow$.   \\
RMSSD & $\downarrow$* & $\downarrow$*   \\
Prc20NN & $\downarrow$. & $\downarrow$   \\
pNN20 & $\downarrow$ & $\downarrow$   \\
HTI & $\downarrow$* & $\downarrow$   \\
TINN &$\downarrow$ & $\downarrow$.   \\ 
HF & $\downarrow$  & $\downarrow$  \\
HFn & $\uparrow$  & $\downarrow$  \\
LnHF & $\downarrow$  & $\downarrow$  \\ 
S & $\downarrow$*  & $\downarrow$*  \\
SD1 & $\downarrow$* & $\downarrow$*  \\
SD2 & $\downarrow$** & $\downarrow$.   \\
SD1SD2 & $\uparrow$ & $\downarrow$   \\
ApEn & $\uparrow$ & $\downarrow$  \\
DFA$\alpha$1 & $\uparrow$  & $\uparrow$ \\
HR & $\uparrow$*  & $\uparrow$  \\ 
Self-reported anxiety (i.e., PAL) &  $\uparrow$  & $\uparrow$** \\ \bottomrule
\end{tabular}
}
\label{tab:gender_table}

\end{table}

\section{Discussion} \label{sec: Discussion}

\subsection{Negative correlation of HRV with self-reported measures}
Through correlation analysis, we found that HRV is negatively associated with self-reported anxiety scores (SPIN and PAL). This suggests that higher anxiety in participants is associated with lower HRV, i.e., the higher the person is anxious, the lower the HRV will be. Furthermore, our analysis indicated that the HRV of female participants has a stronger negative correlation with self-reported anxiety compared to males. This finding suggests that gender plays a significant role in how individuals respond to anxiety-provoking activities. These results imply that HCI practitioners and intervention designers should consider gender-specific factors while developing solutions for SAD.

\subsection{HR, HRV, and PAL across phases and their relationship between SAD and Non-SAD participants}

To answer (i) How does HRV and HR change across phases for all participants? (ii) What are the HRV and HR differences between participants with SAD and those without? (iii) How does self-reports (i.e., PAL) change across phases, regardless of group differences, and how does it differ between participants with SAD and those without?, we hypothesized:  \textbf{Hypothesis 1 (H1):} There will be a significant increase in HR and a decrease in HRV during the Anticipation and Speech activity phases, respectively, compared to the Baseline for all participants. Additionally, HR and HRV will exhibit significant differences during the Reflection phase compared to the Baseline phase for all participants. \textbf{Hypothesis 2 H2):} There will be significant differences in HR and HRV between the SAD and non-SAD participants during the anxious activity. \textbf{Hypothesis 3 (H3):} There will be a significant increase in perceived anxiety levels during the Anticipation and Speech activity phases compared to the Baseline for all participants. Furthermore, perceived anxiety levels during the Reflection phase will significantly differ from those during the Baseline.

In H1, we examined the variations in HRV and HR during the anticipation, speech (activity), and reflection phases, irrespective of the participant groups (SAD vs. non-SAD). We observed notable changes in HRV and HR compared to the baseline in both phases (anticipation and speech). However, the statistical significance of these changes exhibited variations between the phases (see Table \ref{tab:h1}). In the anticipation phase, we found significantly lower SDNN, HTI, TINN, and ApEn and significantly higher HF, HFn, and LnHF values. However, the statistical significance of these differences changed in the speech phase. Similarly, in the speech phase, we found significant decreases in MeanNN, MedianNN, RMSSD, Prc20NN, pNN20, LnHF, S, SD1, SD1SD2, and an increase in HR. Only MedianNN, RMSSD, TINN, S, and SD1 consistently exhibited significant reductions in both anticipation and speech phases compared to the baseline. These findings indicate that while some HRV parameters showed consistent changes across both anticipation and activity phases, others exhibited phase-specific alterations.


In the reflection phase, our observations revealed a mixture of outcomes with notable variations in HRV features. Notably, we observed both significant increases and decreases in various HRV parameters. Specifically, MeanNN, MedianNN, Prc20NN, HF, HFn, and LnHF increased, signifying an augmentation in HRV during this phase. In contrast, HTI, TINN, and ApEn showed a decline. An interesting pattern emerged after excluding HTI, TINN, ApEn, and frequency domain HRV from our analysis:\textit{ HRV features that had exhibited lower values during the anticipation and activity phases demonstrated an increase in the reflection phase}. This pattern suggests that participants may have felt more at ease during the reflection phase compared to the baseline and the preceding phases. This inference aligns with the observed increase in HRV levels and the decrease in HR during this phase, indicating a sense of physiological relaxation and reduced stress among the participants. However, our study methodology could not allow us to assess whether there was also a psychic change (cognition), e.g., self-criticism, being negatively evaluated, and cognitive distortions. Future research can incorporate questions or items on the current level of thinking patterns to gauge these changes. Moreover, a mixed-method study incorporating qualitative methods would be valuable in capturing these nuances of physiological or psychological changes.


In line with Hypothesis 2, we tried to understand the HRV patterns in both the SAD and non-SAD groups. Consistent with previous literature \cite{alvares2013reduced, miranda2014anxiety, tolin2021psychophysiological, garcia2017autonomic, harrewijn2018heart, madison2021social, gaebler2013heart, tolin2021psychophysiological,harrewijn2018heart},  we indeed observed a consistent trend of lower HRV and higher HR in participants with SAD compared non-SAD (see Table \ref{tab:h2}). Nevertheless, RMSSD, S, and SD1 emerged as promising HRV parameters for distinguishing SAD participants from their non-SAD counterparts (see Table \ref{tab:h2}), aligning with the findings of previous studies \cite{alvares2013reduced, garcia2017autonomic,harrewijn2018heart,madison2021social}. These parameters are potentially valuable physiological markers for identifying individuals with SAD. Moreover, our findings on HF are consistent with existing literature \cite{alvares2013reduced,pittig2013heart,gaebler2013heart,tolin2021psychophysiological, harrewijn2018heart}; however, the results in our study were not statistically significant. The correlation analysis revealed that HF was negatively associated with SPIN and PAL during the baseline, anticipation, and reflection phases but showed a positive correlation during the activity phase. This shift suggests that the role of HF may vary across different phases, indicating the need for further investigation.




Furthermore, to understand the change in PAL with phases and between groups, we tested hypothesis 3. Similar to the existing literature~\cite{garcia2017autonomic,mauss2004there}, we found that PAL significantly increased in anticipation and activity  while there was a decrease (insignificant) in the reflection phase. These results imply that the anticipation and performance associated with tasks (speech) induced higher levels of anxiety compared to the baseline as shown in 
 Figure~\ref{fig:self_report_visual}. Furthermore, the analysis revealed a significant difference in PAL between the SAD and non-SAD groups. This finding suggests that individuals with SAD perceive and report higher levels of anxiety than those without SAD. 
Figure~\ref{fig: hrv_sad_vs_non_Sad_figure} shows the changes in the SDNN, RMSSD, SD1, HR, and the PAL across the baseline, anticipation, speech, and reflection.

\subsection{Effect of Gender on HR and HRV}
\begin{figure}
    \begin{subfigure}{0.33\textwidth}
        \includegraphics [width=\textwidth]{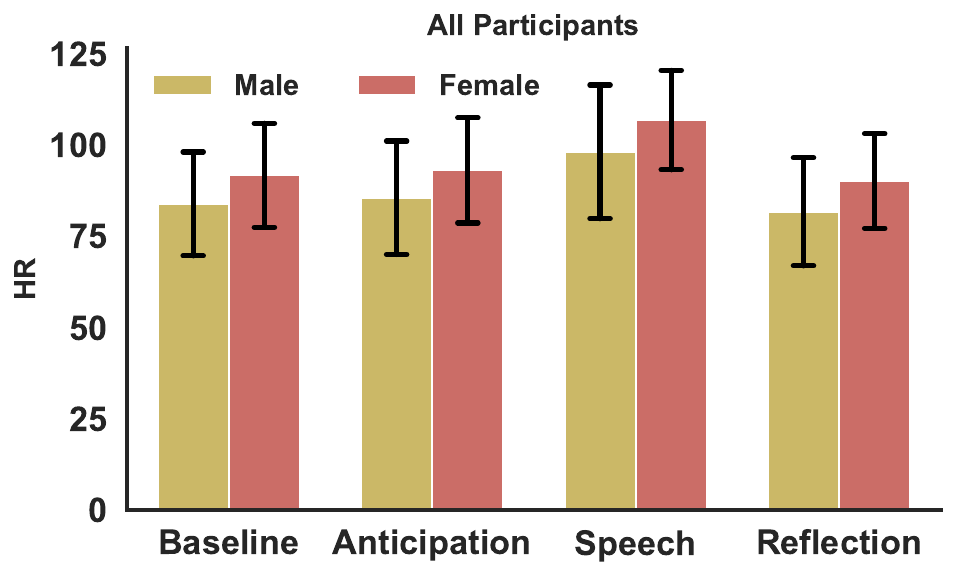}
        \caption{All participants - Male vs. Female}
        \label{fig: ap_mvs_f}
    \end{subfigure} 
        \begin{subfigure}{0.33\textwidth}
        \includegraphics [width=\textwidth]{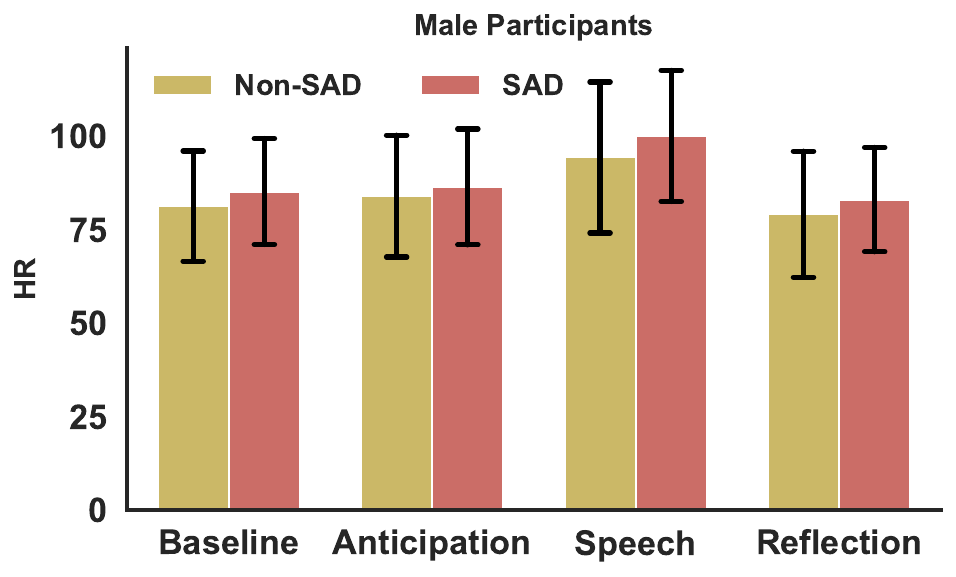}
        \caption{Male participants - SAD vs non-SAD}
        \label{fig: male_sad_vs_non_Sad}
    \end{subfigure} 
    \begin{subfigure}{0.33\textwidth}
        \includegraphics [width=\textwidth]{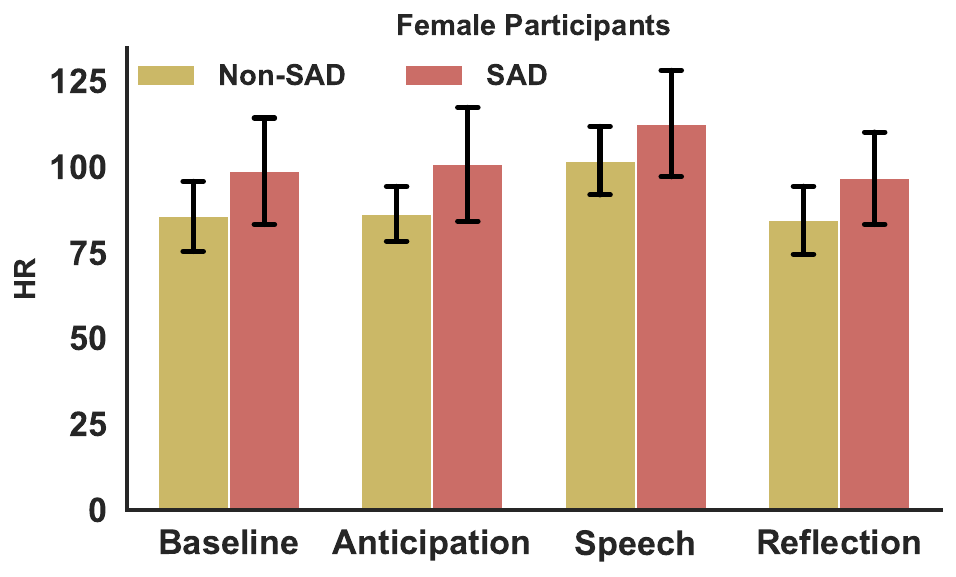}
        \caption{Female participants - SAD vs non-SAD}
        \label{fig: female_sad_vs_non_Sad}
    \end{subfigure} 
    \caption{Differences in HR between different groups. [{\bf Best viewed in color}]}
    \label{fig: male_vs_female_figure}
\end{figure}

In addition to the correlation and hypothesis testing, we explored how HRV and HR vary between males and females. 
The Figure~\ref{fig: male_vs_female_figure} shows the changes in the HR of males and females across the baseline, anticipation, speech, and reflection. In alignment with existing research~\cite{geovanini2020age}, our findings indicated that, as a general trend, females demonstrated a higher HR and lower HRV levels in comparison to males. Furthermore, we discovered that both males and females with SAD displayed lower HRV than their non-SAD counterparts within their respective gender groups~\cite{alvares2013reduced}. 

Our correlation analysis revealed that gender influences HRV and PAL. Based on this, we included gender as a fixed variable in our analysis to understand its effect on HRV while treating groups as a fixed variable. We found that by including gender as a fixed variable, the HRV parameters \textit{TINN} and \textit{SD2} can also serve as physiological markers for SAD. However, the significance level was only 0.1. These gender-specific insights deepen our understanding of HRV patterns and differences related to SAD among males and females. This finding is novel and, to our knowledge, has not been reported in existing literature, warranting further exploration. Moreover, our analysis of gender requires further exploration due to the low number of female participants (33\%) compared to male participants.


\subsection{Physiological markers of SAD}
There is no consensus on which specific HRV features should be prioritized when studying individuals with SAD. To address this gap, our research has focused on examining all time-domain, frequency-domain, and nonlinear indices of HRV during different phases of an anxious activity with respect to the baseline. It is worth noting that the existing literature \cite{miranda2014anxiety, gaebler2013heart, tolin2021psychophysiological,garcia2017autonomic,bailey2019moderating,harrewijn2018heart,madison2021social,pittig2013heart,held2021heart} predominantly focuses on HR, HF, and RMSSD, often overlooking other time-domain and frequency-domain HRV features. Furthermore, only one study has investigated a nonlinear HRV index (only DFA$\alpha$1), and found a significant increase in individuals with SAD \cite{alvares2013reduced}. Although our study yielded similar results for DFA$\alpha$1, the outcomes were not statistically significant. 

Based on our findings and the existing literature, we advocate for a comprehensive exploration of HRV features beyond HF and RMSSD. In particular, we propose that RMSSD, S, and SD1 as the potential physiological markers of SAD. Our study is notable for being the first to identify SD1 (nonlinear HRV), representing short-term RR interval fluctuations, as a potential physiological marker for distinguishing participants with SAD from those without SAD. Additionally, SD1 played a pivotal role in unveiling the dynamic changes in HRV during the different phases of our study. Specifically, they indicate that HRV decreases during the anticipation and activity phases while increasing during the reflection phase compared to the baseline.

Moreover, when considering gender as a factor, TINN and SD2 also emerge as potential physiological markers of SAD (see Table \ref{tab:female}). However, further exploration is needed to validate these physiological markers, as we are the first to propose them as indicators of anxiety disorders with gender as a factor. Examining these features can provide deeper insights into the physiological responses of individuals with SAD and contribute to a more holistic understanding of HRV alterations in gender-specific SAD populations.

\begin{figure}[!h]
        \includegraphics [width=\textwidth]{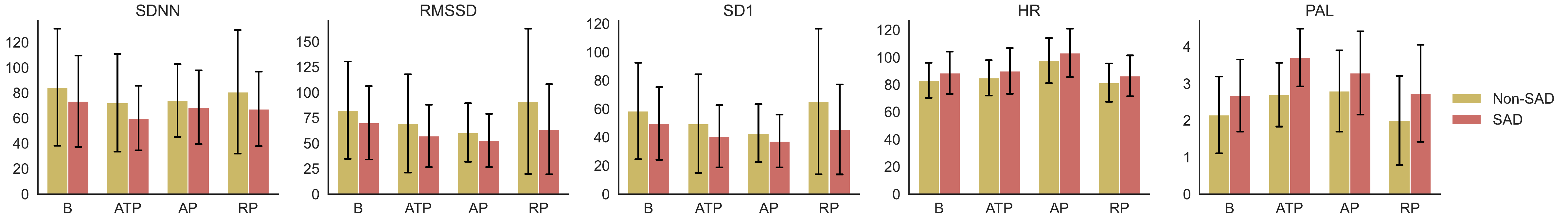}
        \caption{Differences in physiological measures (SDNN, RMSSD, SD1, HR) and 
perceived anxiety level (PAL) across the baseline (B), anticipation (ATP), speech activity (AP), and reflection (RP) phases between SAD and non-SAD participants. [{\bf Best viewed in color}]}
        \label{fig: hrv_sad_vs_non_Sad_figure}
\end{figure}

\subsection{Implications for mental health researchers}

The participants' PAL reports indicate that the anticipation of speech activity provoked higher anxiety levels than the actual performance of the speech itself. The mean PAL scores found during Table~\ref{tab:h3} calculations were as follows: baseline = 2.47, anticipation = 3.31, speech = 3.09, and reflection = 2.45. These scores indicate that the mere announcement about the speech activity induced anxiety in the participants, leading to higher perceived anxiety after the baseline and before the activity commenced. Moreover, the PAL reduced following the completion of the activity, suggesting the participant moved to a relaxed state. These changes in PAL underscore the significance of encompassing both the anticipation and reflection phases rather than solely comparing the anxious activity with the baseline.

\begin{wrapfigure}{r}{0.4\textwidth}  
    \centering
    \includegraphics[scale=0.31]{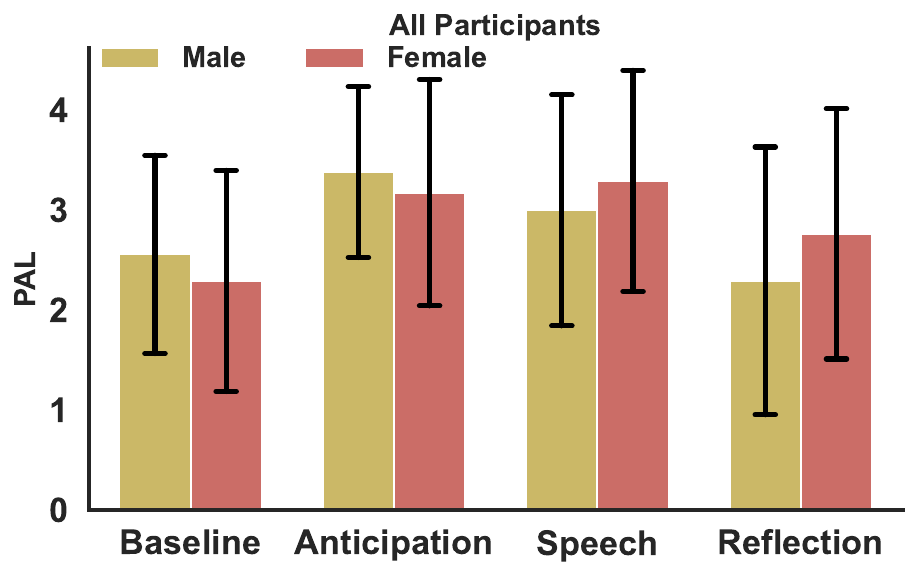}
    \caption{Mean PAL of male and female participants at different phases (Baseline, Anticipation, Speech, and Reflection) of the study. {\bf [Best viewed in color]}}
    \label{fig: PAL_sad_vs_non_Sad_figure}
\end{wrapfigure}

Ideally, any physiological change in the participants should correspond to their PAL. For instance, in continuation with changes observed in the above PAL,  the decrease in HRV should be more prominent during the anticipation phase, followed by the speech phase, and conversely, an opposing trend should be observed during the reflection phase. To this end, we observed changes in MeanNN, MedianNN, Prc20NN, SD1, and SD1SD2 in sync with the PAL. However, it is crucial to note that these findings are irrespective of SAD/non-SAD groups.

The analysis comparing the SAD and non-SAD groups reveals that participants with SAD consistently exhibited reduced HRV during the baseline and all phases compared to the non-SAD group (see Table \ref{tab:h2} and Figure \ref{fig: hrv_sad_vs_non_Sad_figure}). This implies that regardless of anticipation or engaging in the anxious activity, the SAD group consistently demonstrates lower HRV. However, during anticipation and while engaging in the activity, HRV further decreases compared to the baseline state. These results suggest that anxious individuals exhibit lower HRV than non-anxious individuals.

The analysis on the gender reveals that females consistently demonstrated significantly lower heart rate variability (MeanNN, MedianNN, SDNN, RMSSD, HTI, S, SD1, SD2) and higher heart rate levels throughout the study than males (see Table \ref{tab:female}). Nonetheless, it is crucial to highlight that in comparison to the male group, the female group exhibited lower PAL during the baseline and anticipation phases but reported higher PAL during anticipation and reflection (see Figure \ref{fig: PAL_sad_vs_non_Sad_figure}). One possible explanation for this could be linked to higher self-esteem among females (students in our study), as reported in various studies \cite{tamanaifar2023relationship}. Additionally, it is important to note that anxious females exhibited decreased HRV and elevated heart rate compared to non-anxious females. 


\subsubsection{Wearables and data analysis}
Our findings offer following key insights for wearables community and researchers analyzing mental health disorders.

\begin{itemize}    
    \item Firstly, our analysis indicates that RMSSD, SDNN, and SD1 are reliable physiological markers for capturing social anxiety. 
However, our findings are based on data collected with clinical-grade ECG sensors. It is important to note that PPG is easily sensed by smartwatches and serves as a proxy for ECG \cite{jaiswal2023comparative,kinnunen2020feasible, theurl2023smartwatch}. Therefore, PPG should be further explored to validate our findings and extend the use of these physiological markers for SAD detection in real-world settings.

\item  Secondly, our research revealed that incorporating gender as a fixed variable expands the set of physiological markers associated with SAD. Without including gender as a fixed factor, the markers include SDNN, RMSSD, and SD1, whereas, with gender as a fixed factor, the markers also include TINN, S, and SD2 in addition to the previous ones. So, gender should be incorporated while developing SAD prediction models. Moreover, in the current landscape, researchers usually prefer machine learning and deep learning models. However, exploring advanced statistical models incorporating fixed and random effects, such as Logistic regression with mixed effects, could provide the research community with valuable additional insights.
\end{itemize}

\subsection{Contextual relevance of our findings}

Most existing studies on physiological markers for anxiety disorders have been conducted in the Global North \cite{alvares2013reduced, gaebler2013heart, tolin2021psychophysiological, miranda2014anxiety, tamura2013salivary, garcia2017autonomic, madison2021social, held2021heart, harrewijn2018heart, bailey2019moderating}, where healthcare infrastructure is more advanced and mental health services are more accessible. However, the experience and expression of mental disorders are strongly influenced by cultural beliefs, social norms, and systemic stigma, all of which can differ significantly across regions \cite{pendse2019mental}. Factors such as gender roles, social networks, and geopolitics further shape how individuals perceive and respond to mental health challenges \cite{wani2024unrest}.

In contrast, the Global South—particularly countries like India—faces unique challenges, including widespread stigma, limited access to mental health services, and a lack of culturally grounded research \cite{pendse2019mental}. Our study addresses this gap by examining physiological and self-reported markers of SAD within an Indian context. To our knowledge, this is the first study in India to investigate HR and HRV variations during anxiety-inducing activities using wearable sensors.
India, the world’s most populous country, has a highly diverse population spanning multiple cultures, languages, and socioeconomic backgrounds \cite{bose2021integrating}. The participants in our study were students studying at IISER Bhopal, a premier academic institution that draws students from various regions across India. This diversity allowed us to capture various responses across urban and rural settings and genders.

By analyzing both objective physiological data and subjective self-reports from participants with and without SAD, our study provides insights into how anxiety manifests when participants anticipate, perform, and reflect on an anxiety-provoking activity. These findings are essential for designing mental health interventions that are not only effective but are also context-aware. The publicly available dataset generated through this work contributes a rare resource for future research focused on LMICs. It offers a foundation for building localized and inclusive mental health technologies, aligning with the COMPASS vision of computing that is situated, socially responsive, and driven by the realities of place-based challenges.

\section{Limitations and Future work} \label{sec: Limitations} \label{sec: Future_work}

 Our study has several important limitations: (i) Our study primarily involved non-clinical participants (institute students). The SAD and non-SAD categories were grouped based on the SPIN score. Though SPIN is a validated self-reported questionnaire for finding social anxiety in participants, it may not be able to capture the clinically significant SAD. (ii) Data loss: The study was conducted with ninety-nine participants (SAD - 71 and non-SAD - 28), but forty-eight participants (SAD-40 and non-SAD-8) data were dropped due to noise and artifacts. However, ECG data loss is common in cardiovascular research due to noise and artifacts \cite{held2021heart}. The data loss may or may not have affected the results. Future studies should consider implementing thorough skin preparation and enhancing electrode-skin contact to mitigate such issues. These limitations emphasize the need for future research to address these issues to comprehensively understand SAD and its physiological correlates.


Notably, the insights gained from post-discussion interactions with the participants offer valuable considerations for future study design and research endeavors. Specifically: (i) Gender-specific social anxiety: Some male participants, despite having lower SPIN scores, expressed discomfort specifically when interacting with females. Similarly, female participants reported similar experiences. These observations suggest that gender-specific social anxiety may affect how individuals experience and express their anxiety. This aspect could be further explored in future research to better understand the nuances of social anxiety within different social contexts. (ii) Eye Gaze Properties: The research assistant  noted that participants with lower confidence levels tended to avoid eye contact, a commonly used safety behavior among individuals with SAD. This observation introduces an intriguing research question regarding the eye gaze properties of individuals with SAD. Investigating how eye gaze behavior differs between these groups (SAD/ non-SAD) could provide insights into the non-verbal cues associated with social anxiety. These post-discussion insights underscore the multidimensional nature of social anxiety and suggest potential avenues for future research to delve deeper into the intricacies of how individuals experience and manifest social anxiety in various social contexts. Lastly, incorporating a qualitative study design (mixed-method approach) would be valuable in analyzing SAD individuals' physiological and psychological phenomena in a controlled environment.

\section{Conclusion} \label{sec: Conclusion}

Our study used speech as an anxiety-inducing activity and examined HRV and HR before, during, and after this activity. We employed the Shimmer ECG kit, a clinical-grade equipment, to measure ECG signals throughout the study. Further, we extracted HR and HRV features from the collected data corresponding to different phases (anticipation, activity, reflection). Our data analysis was two-fold: we investigated how HR and HRV changed over time irrespective of whether participants had SAD, and we also compared how these HR/HRV features differed between participants with SAD and those without it. The results from our study indicate that all participants experienced a reduction in HRV, with SAD participants exhibiting notably lower HRV than non-SAD participants. Specifically, we found significantly reduced SDNN, RMSSD, and SD1 values in participants with SAD. These findings suggest that these HRV parameters could potentially serve as physiological markers for SAD.


\bibliographystyle{unsrt} 
\bibliography{citations}

@article{sahu2024unveiling,
  title={Unveiling Social Anxiety: Analyzing Acoustic and Linguistic Traits in Impromptu Speech within a Controlled Study},
  author={Sahu, Nilesh Kumar and Yadav, Manjeet and Lone, Haroon R},
  journal={ACM Journal on Computing and Sustainable Societies},
  volume={2},
  number={2},
  pages={1--19},
  year={2024},
  publisher={ACM New York, NY}
}

@article{geovanini2020age,
  title={Age and sex differences in heart rate variability and vagal specific patterns--Baependi heart study},
  author={Geovanini, Glaucylara Reis and Vasques, Enio Rodrigues and de Oliveira Alvim, Rafael and Mill, Jos{\'e} Geraldo and Andre{\~a}o, Rodrigo Varej{\~a}o and Vasques, Bruna Kim and Pereira, Alexandre Costa and Krieger, Jose Eduardo},
  journal={Global heart},
  volume={15},
  number={1},
  year={2020},
  publisher={Ubiquity Press}
}

@inproceedings{pendse2019mental,
  title={Mental health in the global south: challenges and opportunities in HCI for development},
  author={Pendse, Sachin R and Karusala, Naveena and Siddarth, Divya and Gonsalves, Pattie and Mehrotra, Seema and Naslund, John A and Sood, Mamta and Kumar, Neha and Sharma, Amit},
  booktitle={Proceedings of the 2nd ACM SIGCAS Conference on Computing and Sustainable Societies},
  pages={22--36},
  year={2019}
}

@inproceedings{wani2024unrest,
  title={" Unrest and trauma stays with you!": Navigating mental health and professional service-seeking in Kashmir},
  author={Wani, Asra Sakeen and Joshi, Ishika and Nahvi, Nadia Ishfaq and Singh, Pushpendra},
  booktitle={Proceedings of the 2024 CHI Conference on Human Factors in Computing Systems},
  pages={1--17},
  year={2024}
}

@article{bose2021integrating,
  title={Integrating linguistics, social structure, and geography to model genetic diversity within India},
  author={Bose, Aritra and Platt, Daniel E and Parida, Laxmi and Drineas, Petros and Paschou, Peristera},
  journal={Molecular biology and evolution},
  volume={38},
  number={5},
  pages={1809--1819},
  year={2021},
  publisher={Oxford University Press}
}

@article{patel2018lancet,
  title={The Lancet Commission on global mental health and sustainable development},
  author={Patel, Vikram and Saxena, Shekhar and Lund, Crick and Thornicroft, Graham and Baingana, Florence and Bolton, Paul and Chisholm, Dan and Collins, Pamela Y and Cooper, Janice L and Eaton, Julian and others},
  journal={The lancet},
  volume={392},
  number={10157},
  pages={1553--1598},
  year={2018},
  publisher={Elsevier}
}

@article{eshun2009introduction,
  title={Introduction to culture and psychopathology},
  author={Eshun, Sussie and Gurung, Regan AR},
  journal={Culture and mental health: Sociocultural influences, theory, and practice},
  pages={1--17},
  year={2009},
  publisher={Wiley-Blackwell Oxford, UK}
}

@article{fernando2019developing,
  title={Developing mental health services in the global south},
  author={Fernando, Suman},
  journal={International Journal of Mental Health},
  volume={48},
  number={4},
  pages={338--345},
  year={2019},
  publisher={Taylor \& Francis}
}

@article{salamanca2020ethical,
  title={The ethical, social, and cultural dimensions of screening for mental health in children and adolescents of the developing world},
  author={Salamanca-Buentello, Fabio and Seeman, Mary V and Daar, Abdallah S and Upshur, Ross EG},
  journal={PloS one},
  volume={15},
  number={8},
  pages={e0237853},
  year={2020},
  publisher={Public Library of Science San Francisco, CA USA}
}

@book{de2015missed,
  title={Missed opportunities in global health: Identifying new strategies to improve mental health in LMICs},
  author={De Menil, Victoria and Glassman, Amanda},
  year={2015},
  publisher={Center for Global Development}
}

@article{asher2017little,
  title={A little could go a long way: financing for mental healthcare in low-and middle-income countries},
  author={Asher, Laura and De Silva, MJ},
  journal={Epidemiology and psychiatric sciences},
  volume={26},
  number={3},
  pages={248--251},
  year={2017},
  publisher={Cambridge University Press}
}

@article{vaillant2012positive,
  title={Positive mental health: is there a cross-cultural definition?},
  author={Vaillant, George E},
  journal={World Psychiatry},
  volume={11},
  number={2},
  pages={93--99},
  year={2012},
  publisher={Elsevier}
}

@article{stein2008social,
  title={Social anxiety disorder},
  author={Stein, Murray B and Stein, Dan J},
  journal={The lancet},
  volume={371},
  number={9618},
  pages={1115--1125},
  year={2008},
  publisher={Elsevier}
}

@article{ollendick2002developmental,
  title={The developmental psychopathology of social anxiety disorder},
  author={Ollendick, Thomas H and Hirshfeld-Becker, Dina R},
  journal={Biological Psychiatry},
  volume={51},
  number={1},
  pages={44--58},
  year={2002},
  publisher={Elsevier}
}

@article{jaiswal2023comparative,
  title={Comparative Assessment of Smartwatch Photoplethysmography Accuracy},
  author={Jaiswal, Pranay and Sahu, Nilesh Kumar and Lone, Haroon R},
  journal={IEEE Sensors Letters},
  year={2023},
  publisher={IEEE}
}

@article{theurl2023smartwatch,
  title={Smartwatch-derived heart rate variability: a head-to-head comparison with the gold standard in cardiovascular disease},
  author={Theurl, Fabian and Schreinlechner, Michael and Sappler, Nikolay and Toifl, Michael and Dolejsi, Theresa and Hofer, Florian and Massmann, Celine and Steinbring, Christian and Komarek, Silvia and M{\"o}lgg, Kurt and others},
  journal={European Heart Journal-Digital Health},
  volume={4},
  number={3},
  pages={155--164},
  year={2023},
  publisher={Oxford University Press US}
}

@article{kinnunen2020feasible,
  title={Feasible assessment of recovery and cardiovascular health: accuracy of nocturnal HR and HRV assessed via ring PPG in comparison to medical grade ECG},
  author={Kinnunen, Hannu and Rantanen, Aleksi and Kentt{\"a}, Tuomas and Koskim{\"a}ki, Heli},
  journal={Physiological measurement},
  volume={41},
  number={4},
  pages={04NT01},
  year={2020},
  publisher={IOP Publishing}
}

@article{meegahapola2023generalization,
  title={Generalization and Personalization of Mobile Sensing-Based Mood Inference Models: An Analysis of College Students in Eight Countries},
  author={Meegahapola, Lakmal and Droz, William and Kun, Peter and De G{\"o}tzen, Amalia and Nutakki, Chaitanya and Diwakar, Shyam and Correa, Salvador Ruiz and Song, Donglei and Xu, Hao and Bidoglia, Miriam and others},
  journal={Proceedings of the ACM on Interactive, Mobile, Wearable and Ubiquitous Technologies},
  volume={6},
  number={4},
  pages={1--32},
  year={2023},
  publisher={ACM New York, NY, USA}
}

@article{wang2022first,
  title={First-gen lens: Assessing mental health of first-generation students across their first year at college using mobile sensing},
  author={Wang, Weichen and Nepal, Subigya and Huckins, Jeremy F and Hernandez, Lessley and Vojdanovski, Vlado and Mack, Dante and Plomp, Jane and Pillai, Arvind and Obuchi, Mikio and Dasilva, Alex and others},
  journal={Proceedings of the ACM on interactive, mobile, wearable and ubiquitous technologies},
  volume={6},
  number={2},
  pages={1--32},
  year={2022},
  publisher={ACM New York, NY, USA}
}

@article{adler2021identifying,
  title={Identifying mobile sensing indicators of stress-resilience},
  author={Adler, Daniel A and Tseng, Vincent W-S and Qi, Gengmo and Scarpa, Joseph and Sen, Srijan and Choudhury, Tanzeem},
  journal={Proceedings of the ACM on interactive, mobile, wearable and ubiquitous technologies},
  volume={5},
  number={2},
  pages={1--32},
  year={2021},
  publisher={ACM New York, NY, USA}
}

@article{mishra2020evaluating,
  title={Evaluating the reproducibility of physiological stress detection models},
  author={Mishra, Varun and Sen, Sougata and Chen, Grace and Hao, Tian and Rogers, Jeffrey and Chen, Ching-Hua and Kotz, David},
  journal={Proceedings of the ACM on interactive, mobile, wearable and ubiquitous technologies},
  volume={4},
  number={4},
  pages={1--29},
  year={2020},
  publisher={ACM New York, NY, USA}
}

@article{yu2023semi,
  title={Semi-Supervised Learning for Wearable-based Momentary Stress Detection in the Wild},
  author={Yu, Han and Sano, Akane},
  journal={Proceedings of the ACM on Interactive, Mobile, Wearable and Ubiquitous Technologies},
  volume={7},
  number={2},
  pages={1--23},
  year={2023},
  publisher={ACM New York, NY, USA}
}

@article{wang2023detecting,
  title={Detecting Social Contexts from Mobile Sensing Indicators in Virtual Interactions with Socially Anxious Individuals},
  author={Wang, Zhiyuan and Larrazabal, Maria A and Rucker, Mark and Toner, Emma R and Daniel, Katharine E and Kumar, Shashwat and Boukhechba, Mehdi and Teachman, Bethany A and Barnes, Laura E},
  journal={Proceedings of the ACM on Interactive, Mobile, Wearable and Ubiquitous Technologies},
  volume={7},
  number={3},
  pages={1--26},
  year={2023},
  publisher={ACM New York, NY, USA}
}

@inproceedings{akbar2019email,
  title={Email makes you sweat: Examining email interruptions and stress using thermal imaging},
  author={Akbar, Fatema and Bayraktaroglu, Ayse Elvan and Buddharaju, Pradeep and Da Cunha Silva, Dennis Rodrigo and Gao, Ge and Grover, Ted and Gutierrez-Osuna, Ricardo and Jones, Nathan Cooper and Mark, Gloria and Pavlidis, Ioannis and others},
  booktitle={Proceedings of the 2019 CHI Conference on Human Factors in Computing Systems},
  pages={1--14},
  year={2019}
}

@article{cheng2022heart,
  title={Heart rate variability in patients with anxiety disorders: A systematic review and meta-analysis},
  author={Cheng, Ying-Chih and Su, Min-I and Liu, Cheng-Wei and Huang, Yu-Chen and Huang, Wei-Lieh},
  journal={Psychiatry and Clinical Neurosciences},
  volume={76},
  number={7},
  pages={292--302},
  year={2022},
  publisher={Wiley Online Library}
}

@inproceedings{paredes2018fast,
  title={Fast \& furious: detecting stress with a car steering wheel},
  author={Paredes, Pablo E and Ordonez, Francisco and Ju, Wendy and Landay, James A},
  booktitle={Proceedings of the 2018 CHI conference on human factors in computing systems},
  pages={1--12},
  year={2018}
}

@article{bates2014fitting,
  title={Fitting linear mixed-effects models using lme4},
  author={Bates, Douglas and M{\"a}chler, Martin and Bolker, Ben and Walker, Steve},
  journal={arXiv preprint arXiv:1406.5823},
  year={2014}
}

@misc{bristol_multilevel,
  title = {What are multilevel models and why should I use them?},
  note={\url{https://www.bristol.ac.uk/cmm/learning/multilevel-models/what-why.html}, accessed date: 2024-12-07},
}

@article{mauss2004there,
  title={Is there less to social anxiety than meets the eye? Emotion experience, expression, and bodily responding},
  author={Mauss, Iris and Wilhelm, Frank and Gross, James},
  journal={Cognition and emotion},
  volume={18},
  number={5},
  pages={631--642},
  year={2004},
  publisher={Taylor \& Francis}
}

@article{held2021heart,
  title={Heart rate variability change during a stressful cognitive task in individuals with anxiety and control participants},
  author={Held, Judith and V{\^\i}sl{\u{a}}, Andreea and Wolfer, Christine and Messerli-B{\"u}rgy, Nadine and Fl{\"u}ckiger, Christoph},
  journal={BMC psychology},
  volume={9},
  number={1},
  pages={1--8},
  year={2021},
  publisher={BioMed Central}
}

@article{madison2021social,
  title={Social anxiety symptoms, heart rate variability, and vocal emotion recognition in women: evidence for parasympathetically-mediated positivity bias},
  author={Madison, Annelise and Vasey, Michael and Emery, Charles F and Kiecolt-Glaser, Janice K},
  journal={Anxiety, Stress, \& Coping},
  volume={34},
  number={3},
  pages={243--257},
  year={2021},
  publisher={Taylor \& Francis}
}

@article{bailey2019moderating,
  title={Moderating effects of the valence of social interaction on the dysfunctional consequences of perseverative cognition: an ecological study in major depression and social anxiety disorder},
  author={Bailey, Tamara and Shahabi, L and Tarvainen, M and Shapiro, D and Ottaviani, C},
  journal={Anxiety, Stress, \& Coping},
  volume={32},
  number={2},
  pages={179--195},
  year={2019},
  publisher={Taylor \& Francis}
}

@article{tamura2013salivary,
  title={Salivary alpha-amylase and cortisol responsiveness following electrical stimulation stress in patients with the generalized type of social anxiety disorder},
  author={Tamura, A and Maruyama, Y and Ishitobi, Y and Kawano, A and Ando, T and Ikeda, R and Inoue, A and Imanaga, J and Okamoto, S and Kanehisa, M and others},
  journal={Pharmacopsychiatry},
  volume={46},
  number={07},
  pages={225--260},
  year={2013},
  publisher={{\copyright} Georg Thieme Verlag KG}
}

@article{tolin2021psychophysiological,
  title={Psychophysiological assessment of stress reactivity and recovery in anxiety disorders},
  author={Tolin, David F and Lee, Eric and Levy, Hannah C and Das, Akanksha and Mammo, Liya and Katz, Benjamin W and Diefenbach, Gretchen J},
  journal={Journal of Anxiety Disorders},
  volume={82},
  pages={102426},
  year={2021},
  publisher={Elsevier}
}

@article{harrewijn2018heart,
  title={Heart rate variability as candidate endophenotype of social anxiety: A two-generation family study},
  author={Harrewijn, A and Van der Molen, MJW and Verkuil, B and Sweijen, SW and Houwing-Duistermaat, JJ and Westenberg, PM},
  journal={Journal of Affective Disorders},
  volume={237},
  pages={47--55},
  year={2018},
  publisher={Elsevier}
}

@inproceedings{lecamwasam2023investigating,
  title={Investigating the Physiological and Psychological Effect of an Interactive Musical Interface for Stress and Anxiety Reduction},
  author={Lecamwasam, Kimaya and Gutierrez Arango, Samantha and Singh, Nikhil and Elhaouij, Neska and Addae, Max and Picard, Rosalind},
  booktitle={Extended Abstracts of the 2023 CHI Conference on Human Factors in Computing Systems},
  pages={1--9},
  year={2023}
}

@inproceedings{xue2022understanding,
  title={Understanding How Group Workers Reflect on Organizational Stress with a Shared, Anonymous Heart Rate Variability Data Visualization},
  author={Xue, Mengru and Liang, Rong-Hao and Hu, Jun and Yu, Bin and Feijs, Loe},
  booktitle={CHI Conference on Human Factors in Computing Systems Extended Abstracts},
  pages={1--7},
  year={2022}
}

@book{leyland2001multilevel,
  title={Multilevel modelling of health statistics},
  author={Leyland, Alastair H and Goldstein, Harvey},
  year={2001},
  publisher={Wiley}
}

@book{leyland2020multilevel,
  title={Multilevel modelling for public health and health services research: health in context},
  author={Leyland, Alastair H and Groenewegen, Peter P},
  year={2020},
  publisher={Springer Nature}
}

@inproceedings{reyero2022heart,
  title={Heart rate variability for non-intrusive cybersickness detection},
  author={Reyero Lobo, Paula and Perez, Pablo},
  booktitle={ACM International Conference on Interactive Media Experiences},
  pages={221--228},
  year={2022}
}

@inproceedings{urrestilla2020measuring,
  title={Measuring cognitive load: Heart-rate variability and pupillometry assessment},
  author={Urrestilla, Nerea and St-Onge, David},
  booktitle={Companion Publication of the 2020 International Conference on Multimodal Interaction},
  pages={405--410},
  year={2020}
}

@article{shaffer2017overview,
  title={An overview of heart rate variability metrics and norms},
  author={Shaffer, Fred and Ginsberg, Jay P},
  journal={Frontiers in public health},
  pages={258},
  year={2017},
  publisher={Frontiers}
}

@book{lauria2016primary,
  title={The primary care toolkit for anxiety and related disorders: Quick, practical solutions for assessment and management},
  author={Lauria-Horner, Bianca},
  year={2016},
  publisher={Brush Education}
}

@article{makowski2021neurokit2,
  title={NeuroKit2: A Python toolbox for neurophysiological signal processing},
  author={Makowski, Dominique and Pham, Tam and Lau, Zen J and Brammer, Jan C and Lespinasse, Fran{\c{c}}ois and Pham, Hung and Sch{\"o}lzel, Christopher and Chen, SH Annabel},
  journal={Behavior research methods},
  pages={1--8},
  year={2021},
  publisher={Springer}
}

@article{chukwujekwu2018validation,
  title={Validation of the social phobia inventory (Spin) in Nigeria},
  author={Chukwujekwu, DC and Olose, EO},
  journal={Journal of Psychiatry and Psychiatric Disorders},
  volume={2},
  number={2},
  pages={49--54},
  year={2018},
  publisher={Fortune Journals}
}

@article{grossman2001gender,
  title={Gender differences in psychophysiological responses to speech stress among older social phobics: congruence and incongruence between self-evaluative and cardiovascular reactions},
  author={Grossman, Paul and Wilhelm, Frank H and Kawachi, Ichiro and Sparrow, David},
  journal={Psychosomatic Medicine},
  volume={63},
  number={5},
  pages={765--777},
  year={2001},
  publisher={LWW}
}

@inproceedings{hamilton2002open,
  title={Open source {ECG} analysis},
  author={Hamilton, Pat},
  booktitle={Computers in cardiology},
  pages={101--104},
  year={2002},
  organization={IEEE}
}

@article{mauss2003autonomic,
  title={Autonomic recovery and habituation in social anxiety},
  author={Mauss, Iris B and Wilhelm, Frank H and Gross, James J},
  journal={Psychophysiology},
  volume={40},
  number={4},
  pages={648--653},
  year={2003},
  publisher={Wiley Online Library}
}

@article{alvares2013reduced,
  title={Reduced heart rate variability in social anxiety disorder: associations with gender and symptom severity},
  author={Alvares, Gail A and Quintana, Daniel S and Kemp, Andrew H and Van Zwieten, Anita and Balleine, Bernard W and Hickie, Ian B and Guastella, Adam J},
  journal={PloS one},
  volume={8},
  number={7},
  pages={e70468},
  year={2013},
  publisher={Public Library of Science San Francisco, USA}
}

@article{pittig2013heart,
  title={Heart rate and heart rate variability in panic, social anxiety, obsessive--compulsive, and generalized anxiety disorders at baseline and in response to relaxation and hyperventilation},
  author={Pittig, Andre and Arch, Joanna J and Lam, Chi WR and Craske, Michelle G},
  journal={International journal of psychophysiology},
  volume={87},
  number={1},
  pages={19--27},
  year={2013},
  publisher={Elsevier}
}

@article{tamanaifar2023relationship,
  title={The relationship of emotional intelligence, self concept and self esteem to academic achivenment},
  author={Tamanaifar, Mohammad Reza and Sedighi Arfai, Fariborz and Salami Mohammad Abadi, Fatemeh},
  journal={Quarterly journal of research and planning in higher education},
  volume={16},
  number={2},
  pages={99--113},
  year={2023},
  publisher={Institute for Research and Planning in Higher Education}
}

@misc{havard_hrv,
  title = {Heart rate variability: How it might indicate well-being},
  note={\url{https://www.health.harvard.edu/blog/heart-rate-variability-new-way-track-well-2017112212789}, accessed date: 2024-12-07},
}

@article{gaebler2013heart,
  title={Heart rate variability and its neural correlates during emotional face processing in social anxiety disorder},
  author={Gaebler, Michael and Daniels, Judith K and Lamke, Jan-Peter and Fydrich, Thomas and Walter, Henrik},
  journal={Biological psychology},
  volume={94},
  number={2},
  pages={319--330},
  year={2013},
  publisher={Elsevier}
}

@article{garcia2017autonomic,
  title={Autonomic markers associated with generalized social phobia symptoms: heart rate variability and salivary alpha-amylase},
  author={Garcia-Rubio, Maria J and Espin, Laura and Hidalgo, Vanesa and Salvador, Alicia and Gomez-Amor, Jesus},
  journal={Stress},
  volume={20},
  number={1},
  pages={61--68},
  year={2017},
  publisher={Taylor \& Francis}
}

@article{salekin2018weakly,
  title={A weakly supervised learning framework for detecting social anxiety and depression},
  author={Salekin, Asif and Eberle, Jeremy W and Glenn, Jeffrey J and Teachman, Bethany A and Stankovic, John A},
  journal={Proceedings of the ACM on interactive, mobile, wearable and ubiquitous technologies},
  volume={2},
  number={2},
  pages={1--26},
  year={2018},
  publisher={ACM New York, NY, USA}
}

@article{connor2000psychometric,
  title={Psychometric properties of the Social Phobia Inventory (SPIN): New self-rating scale},
  author={Connor, Kathryn M and Davidson, Jonathan RT and Churchill, L Erik and Sherwood, Andrew and Weisler, Richard H and Foa, Edna},
  journal={The British Journal of Psychiatry},
  volume={176},
  number={4},
  pages={379--386},
  year={2000},
  publisher={Cambridge University Press}
}

@article{rashid2020predicting,
  title={Predicting subjective measures of social anxiety from sparsely collected mobile sensor data},
  author={Rashid, Haroon and Mendu, Sanjana and Daniel, Katharine E and Beltzer, Miranda L and Teachman, Bethany A and Boukhechba, Mehdi and Barnes, Laura E},
  journal={Proceedings of the ACM on Interactive, Mobile, Wearable and Ubiquitous Technologies},
  volume={4},
  number={3},
  pages={1--24},
  year={2020},
  publisher={ACM New York, NY, USA}
}

@inproceedings{miranda2014anxiety,
  title={Anxiety detection using wearable monitoring},
  author={Miranda, Dari{\'e}n and Calder{\'o}n, Marco and Favela, Jesus},
  booktitle={Proceedings of the 5th Mexican conference on human-computer interaction},
  pages={34--41},
  year={2014}
}

@article{asher2020dating,
  title={Dating with social anxiety: An empirical examination of momentary anxiety and desire for future interaction},
  author={Asher, Maya and Aderka, Idan M},
  journal={Clinical Psychological Science},
  volume={8},
  number={1},
  pages={99--110},
  year={2020},
  publisher={Sage Publications Sage CA: Los Angeles, CA}
}

@article{asher2020out,
  title={Out of sync: nonverbal synchrony in social anxiety disorder},
  author={Asher, Maya and Kauffmann, Amitay and Aderka, Idan M},
  journal={Clinical Psychological Science},
  volume={8},
  number={2},
  pages={280--294},
  year={2020},
  publisher={Sage Publications Sage CA: Los Angeles, CA}
}

\end{document}